\documentclass[11pt,a4paper]{article}
\usepackage{jcappub}
\usepackage{amsmath,amssymb}
\usepackage{epsfig,graphicx}

\begin{document}

\tolerance=5000

\title{Cosmological attractor inflation from the RG-improved Higgs sector of finite gauge
theory}

\author[a]{Emilio~Elizalde,}
\affiliation[a]{Instituto de Ciencias del Espacio (ICE/CSIC) \, and
\\
Institut d'Estudis Espacials de Catalunya (IEEC), Campus UAB, Carrer de Can Magrans, s/n 
08193 Cerdanyola del Vall\`{e}s (Barcelona), Spain} \emailAdd{elizalde@ieec.uab.es}
\author[a,b]{Sergei~D.~Odintsov,}
\affiliation[b]{Instituci\'o Catalana de Recerca i Estudis Avancats
(ICREA), Barcelona, Spain}\emailAdd{odintsov@ieec.uab.es}
\author[c]{Ekaterina~O.~Pozdeeva,}
\affiliation[c]{Skobeltsyn Institute of Nuclear Physics,  Lomonosov
Moscow State University, Leninskie Gory 1, 119991, Moscow, Russia}
 \emailAdd{pozdeeva@www-hep.sinp.msu.ru}
\author[c]{Sergey Yu. Vernov}
\emailAdd{svernov@theory.sinp.msu.ru}

%\date{ \ }

\abstract{
The possibility to construct an inflationary scenario for
renormalization-group improved
potentials corresponding to the Higgs sector of finite gauge models is
investigated. Taking into account quantum corrections to the
renormalization-group potential which sums all leading logs of
perturbation theory is essential
for a successful realization of the inflationary scenario, with very reasonable
parameter values. The
inflationary models thus obtained are seen to be in good agreement with
the most recent and accurate
observational data. More specifically, the values of the relevant inflationary
parameters, $n_\mathrm{s}$ and $r$,
are close to the corresponding ones in the $R^2$ and Higgs-driven inflation
scenarios. It is shown that the model here constructed and Higgs-driven inflation
belong to the same class of cosmological attractors.
}


\arxivnumber{1509.08817}

\maketitle

\section{Introduction}
The existence of an extremely short
stage of accelerated expansion in the very early Universe
(inflation) provides a simple explanation of astronomical data, including the fact that, at cosmological distances, the Universe is approximately isotropic, homogeneous, and spatially flat. Models of cosmic inflation yield accurate quantitative predictions for a number of observable quantities~\cite{Lindebook,19,Inflation_review}. It is known that scalar fields play an essential role in the current description of the evolution of the Universe at a very early epoch~\cite{Linde:1981mu,Albrecht:1982wi,Salopek,inflation2}.  Modified gravity inflationary models~\cite{NO2011,Faulkner:2006ub,Sebastiani:2015kfa}, for example, $R^2$ inflation~\cite{Starobinsky:1979ty,Mukhanov:1981xt}, which can be considered as generic General Relativity models with additional scalar fields, are quite popular as well.

The confirmed discovery of the Higgs boson at the Large Hadron Collider (CERN) has initiated an intense research activity with the aim to understand the cosmological implications of this truly fundamental scalar field.
A really crucial issue in this respect is the possibility to describe inflation
using particle physics~\cite{nonmin-quant,Cervantes-Cota1995,Lyth:1998xn}, as
the Standard Model of elementary
particles itself~\cite{BezShapo,HiggsInflation,DeSimone:2008ei,Lerner,HiggsRG}, or either supersymmetric
models~\cite{SUSEinflation,Carrasco:2015pla,Pallis:2013yda}, or even non-supersymmetric grand unified
theories (GUTs)~\cite{Albrecht:1982wi,nonmin-quant,GUT_Inflation}.

Quantum field theory in curved space-time is an unavoidable, extremely important concept at the very early universe, where curvature is large and typical energies are very high.
In this situation, quantum effects necessarily modify the gravitational action and may cause inflation to occur, as well as other interesting astrophysical phenomena. For instance, it is well-known that quantum field theories in curved space-time lead to curvature-induced phase transitions
(for a detailed description, see~\cite{BOS,BO1985,Elizalde:1993ee}). More precisely, these curvature-induced phase transitions may simply and naturally explain the origin of
inflation itself, especially, by taking into account the point of view that the inflaton
might be nothing else but the Higgs field.

As discussed in~\cite{BOS,BO1985}, the most completed description of curvature
induced phase transitions can be given in terms of the summation of all
leading-logs of quantum field theory, when one considers this issue within
the renormalization-group improved effective potential perspective (see~\cite{Elizalde:1993ee,Elizalde:1994im}). Indeed, in this case, the
corresponding RG-improved effective potential goes far beyond the one-loop
approximation. Let us stress, once more, that these phase transitions are in fact
very important in the early universe. In particular, a large number of models
of the inflationary universe~\cite{Inflation_review} are based on
 first-order phase transitions, which took place during the
 reheating phase of the Universe, in the epoch when the grand unification
 description is applied~\cite{Albrecht:1982wi}.
Hence, in the absence of a consistent theory of quantum gravity, all classical
 theories should better be treated as quantum field theories in curved
 space-time, as extensively discussed
 in~\cite{BOS,Elizalde:1993ee}. In fact, some recent results by the Planck
 collaboration~\cite{Planck2013,Planck2015} seem to point towards the GUT
 scale, what is quite a remarkable connection between GUTs and inflation.
And, in this context, GUTs ought to be treated as quantum field theories in the curved
 space-time corresponding to the very early universe. The calculation of beta-functions which was done in previous works (as in the book by~\cite{BOS}) is based on the use of dimensional regularization and, hence, it does not
depend on any explicit cut-off choice. We
consider that Quantum Gravity effects are less relevant in this approach,
since we work below the Planck scale.

As a consequence, it is natural to start with the issue of the
renormalization-group improved effective potential for an arbitrary renormalizable
massless gauge theory in curved space-time~\cite{Elizalde:1993ee}.
Note that it is enough to work in the linear curvature approximation,
because these linear curvature terms are expected to give
dominant contribution in the discussion of the inflationary effective
potential corresponding to GUTs.  Observe also that we work with Higgs-like inflation where
the scalar potential receives quantum corrections. In this case, the linear curvature
term is the leading one, as compared with the $R^2$ term, which is also induced
by quantum corrections. Indeed, it is known that Starobinsky's $R^2$ inflation is
classically equivalent to Higgs inflation, because both models generate a specific exponential
potential in the Einstein frame. Hence, the model under discussion here, constrained by a scalar potential
and a linear curvature term, is also classically equivalent to a
specific $F(R)$ gravity theory.  By generalizing the
flat space-time Coleman--Weinberg potential~\cite{Coleman}, the explicit form of the
renormalization group (RG) improved effective potentials in curved space finite gauge
models were obtained in~\cite{Elizalde:1994im}. The occurrence of curvature-induced
phase transitions was also studied there in detail.

It is well-known that in high energy physics there exist so-called
finite (gauge) theories. There are two classes of such theories.
The first class are the theories in which, by some reason, the corresponding coupling
constants are not renormalized up to some loop. For instance, those could
be one-loop finite theories. This simply means that the corresponding one-loop
beta-functions are zero but, at the next order in the loop expansion the beta-functions
do receive corrections. The second, and more important, class of finite theories is
usually a consequence of supersymmetry. Indeed, a few supersymmetric theories have been proven to
be finite to all orders of perturbation theory ($N=4$ super Yang--Mills theory~\cite{Stelle} being a well known example). However, it is a fact that supersymmetric theories
are not the only ones which can be finite at the one- or two-loop levels;
different GUTs have been proposed which turn out to yield finite models,
too. Asymptotically finite GUTs~\cite{Ermushev8} are generalizations of the
concept of a finite theory, in which the zero charge problem is absent.
Indeed, both in the UV and in the IR limits the effective coupling constants
tend to some constant values (corresponding to finite phases).
However, all the above remarks are about finite theories in flat space.
When we consider such finite theories in curved spacetime the situation
changes qualitatively. The point is that even the finite gauge theories to all loops are
not finite in curved spacetime. There are two sectors where the theory
ceases to be finite. First of all, in order to make the theory
multiplicatively-renormalizable in curved spacetime one has to add to the
matter Lagrangian the so-called Lagrangian of the external gravitational field
(vacuum polarization terms) (see the book~\cite{BOS}). There appear
corresponding couplings and beta-functions for this external field
Lagrangian. The leading contribution to these beta-functions is proportional to
a number of fields (it maybe read off from the coefficients of the conformal
anomaly) and is defined by the structure of the free matter
Lagrangian. There is no way to make these external couplings
beta-functions to be zero. The second sector where non-zero one-loop beta
functions appear is the non-minimal coupling of scalar field with curvature.
Again, the corresponding beta-function cannot be made zero for arbitrary
values of the non-minimal coupling $\xi$. Hence, even being a theory finite in flat
space-time, it becomes non-finite (or finite only over part of the coupling
constants) in curved spacetime. Due to multiplicative renormalizability of
such theory, one can apply the standard methods of renormalization group
for such partly finite gauge theory in order to get the effective
potential. In this way, we expect to find a high-energy physics motivation
for the class of exponential potentials to describe the inflationary
universe. (Note that one of the main problems of inflationary cosmology is the
fact that many inflationary potentials are taken ad hoc, without any
physical justification for the corresponding choice). Furthermore,
it maybe expected that some flat-space finite SUSY theories (like $N=4$
super Yang--Mills theory) are relevant precisely at the very early universe when
inflation starts.

Very recent observations~\cite{WMAP,Planck2013,Planck2015} (see also~\cite{Seljak}) result in important restrictions on existing inflationary models. The temperature data of the Planck full mission and the first release of polarization data on large angular scales~\cite{Planck2015}, constraint the
spectral index of curvature perturbations to be
\begin{equation}\label{nsdata}
    n_\mathrm{s}=0.968 \pm 0.006 \quad (68\%\, c.l.),
\end{equation}
and the upper bound on the tensor-to-scalar ratio, as
\begin{equation}\label{rdata}
    r < 0.11 \quad (95\%\, c.l.).
\end{equation}

The high-precision measurements performed by the Planck survey show that non-Gaussian perturbations are quite small, what makes single-field inflationary models become more realistic.
At the same time, the predictions of the simplest inflationary models with a minimally coupled scalar field lead to rather large values of the tensor-to-scalar ratio of the density perturbations $r$, and are therefore ruled out by recent Planck data~\cite{Planck2013,Planck2015}.
Such inflationary scenarios can be improved by adding a tiny non-minimal coupling of the inflaton field to gravity~\cite{GB2013,KL2013}. This is not so artificial as it might seem at first sight, since one should note that generic quantum corrections to the action of the scalar field minimally coupled to gravity do include a non-minimally coupling term~\cite{Chernikov}, proportional to $R\phi^2$ and, as is clear, quantum corrections must definitely be taken into account at the inflationary scales. The non-minimal $R\phi^2$ term is always induced by quantum corrections and its presence assures the multiplicative renormalizability of scalar theory in
 curved space-time. Inflationary models with the Ricci scalar multiplied by a function of the scalar field are being intensively studied in
cosmology~\cite{GB2013,KL2013,Salopek,nonmin-infl,OdintsovNOnmin,nonmin-quant,Cerioni,Cooper:1982du,Kaiser,
Cervantes-Cota1995,KKhT,BezShapo, HiggsInflation, DeSimone:2008ei,Lerner,HiggsRG,EOPV2014,Inagaki:2014wva,LindeKallosh,
Kallosh:2014rha,LindeKallosh1,LindeKallosh2}.

There are several inflationary models which are described by the renormalization group improved effective action~\cite{DeSimone:2008ei,HiggsRG,EOPV2014,Inagaki:2014wva,rgi,Inagaki:2015eza}.
In our paper~\cite{EOPV2014}, we showed that both for scalar electrodynamics and for the $SU(5)$ RG-improved models,\footnote{The corresponding potentials had been proposed in~\cite{Elizalde:1993ee,Elizalde:1994im}.} inflationary scenarios are possible and they are in good agreement with the most recent astronomical data~\cite{Planck2013,Seljak}, provided some reasonable values are taken for the parameters.
In~\cite{EOPV2014} we checked the possibility to construct inflationary models using RG-improved effective potentials, considering inflation based on de Sitter solutions, their instability providing a graceful exit from inflation.
In the present paper we make a step forward and show that RG-corrections can {\it generate} a Hilbert--Einstein term in the action. In other words, we do not need to include this term by hand, as is done in Higgs-driven inflation~\cite{BezShapo,HiggsInflation,DeSimone:2008ei,Lerner,HiggsRG}, since we get it naturally, as part of the quantum corrections to the induced gravity term.

There are many different models of inflation, some of them giving very close predictions to each other, for the observable inflationary parameters. For instance, the two different models, $R^2$ Starobinsky inflation~\cite{Starobinsky:1979ty} and
Higgs-driven inflation~\cite{BezShapo,HiggsInflation}, yield very similar values for the spectral index of the curvature perturbations $n_\mathrm{s}$ and for
the tensor-to-scalar ratio of the density perturbations $r$, respectively. In both models~\cite{BezrukovR2Higgs}, one gets approximately
\begin{equation}
\label{nsrN}
n_\mathrm{s}\simeq 1-\frac{2}{N_e},\qquad r\simeq \frac{12}{N_e^2},
\end{equation}
where $N_e$ is the number of e-foldings during inflation. This number must be matched with the appropriate normalization of the data set and with the cosmic history. For the standard interval $50\leqslant N_e\leqslant65$ formula (\ref{nsrN}) gives suitable values of $n_\mathrm{s}$ and $r$. In recent papers~\cite{LindeKallosh,Kallosh:2014rha,LindeKallosh1,LindeKallosh2,Muh,Roest:2013fha,Binetruy:2014zya,Ventury2015,Pieroni2015} it has been shown that there are several classes of inflationary models such that all models within a given class predict the same values of $n_\mathrm{s}$ and $r$ in the leading approximation in $1/N_e$.

In the present paper we consider models with non-minimally coupled scalar field and finite gauge RG-improved potentials. We show that for the finite gauge models, the inflationary scenario can be realized without having an exact de Sitter solution, but indeed from a quasi-de Sitter solution with a slowly changing Hubble parameter. We calculate the corresponding spectral index of curvature perturbations, the tensor-to-scalar ratio of density perturbations, and the running of the spectral index in the new model and show that an inflationary model with this potential is truly compatible with the most recent cosmological data. We will see that our model here  yields the same inflationary parameters as the $R^2$ and Higgs-driven inflationary models and does belong to the same class of cosmological attractors as Higgs-driven inflation and $R^2$ Starobinsky inflation. We also show that the cosmological attractor method is useful to get inflationary parameters for the model considered.

The paper is organized as follows. In Sect.~2 we recall the basic formulae to explore inflationary models with non-minimally coupled scalar fields. Sect.~3 is devoted to the construction of inflationary models based on the RG-improved Higgs sector of the finite gauge model. In Sect.~4 a comparison is carried out of the inflationary scenarios here obtained with the general class of inflationary models known as $\alpha$--attractors. In Sect.~5 we compare the model considered with the $R^2$~inflationary scenario. Finally, the last section is devoted to conclusions and prospects for future research.

\section{Inflationary models with non-minimal coupling}

Let us consider a gravity model with a non-minimally coupled scalar field, described by the action
\begin{equation}
\label{action} S=\int d^4 x \sqrt{-g}\left[
U(\phi)R-\frac12g^{\mu\nu}\phi_{,\mu}\phi_{,\nu}-V(\phi)\right],
\end{equation}
where $U(\phi)$ and $V(\phi)$ are differentiable functions of the
scalar field $\phi$, $g$ is the determinant of the metric tensor
$g_{\mu\nu}$, and $R$ is the scalar curvature.

In a spatially flat FLRW universe, with the interval
\begin{equation*}
ds^2={}-dt^2+a^2(t)\left(dx_1^2+dx_2^2+dx_3^2\right),
\end{equation*}
the Friedmann equations, derived by variation of the
action~(\ref{action}), have the following form~\cite{KTV2011}:
\begin{equation}
\label{Fr1} 6UH^2+6\dot U H=\frac{1}{2}\dot\phi^2+V\,,
\end{equation}
\begin{equation}
\label{Fr2} 2U\left(2\dot H+3H^2\right)+4\dot U H+2\ddot U={}-\frac{1}{2}\dot\phi^2+V\,,
\end{equation}
where the Hubble parameter is the logarithmic derivative of the
scale factor: $H=\dot a/a$ and differentiation with respect to
time $t$ is denoted by a dot.
Variation of the action
(\ref{action}) with respect to $\phi$ yields
\begin{equation}
\label{Fieldequ} \ddot \phi+3H\dot\phi+V^{\prime}=6\left(\dot H
+2H^2\right)U^{\prime}\,,
\end{equation}
where the prime denotes derivative with respect to the  scalar field $\phi$.

The standard way to calculate the parameters of inflation is to perform a conformal transformation and consider the model in the Einstein frame (see, for instance~\cite{DeSimone:2008ei}).
There is an ongoing discussion about the physical equivalence of these two frames~\cite{frames}, in special on the equivalence of the corresponding quantum theories~\cite{Lerner,Steinwachs:2011zs,Kamenshchik:2014waa} (at the classical level the issue seems to be clear by now). In our paper we consider the Jordan frame to be the physical one. For this reason, we calculate quantum corrections in the Jordan frame. The calculation of the $\beta$ functions in the Einstein frame potential may, therefore, yield a different result. On the other hand, once the quantum corrections have been obtained, one can used the Einstein frame to get the inflationary parameters. This is possible because of the quasi de Sitter expansion occurring during inflation. Indeed, it has been shown~\cite{Kaiser}, that in the case of quasi de Sitter expansion there is no difference between the inflationary parameters calculated either in the Jordan frame directly, or in the Einstein frame, after the corresponding conformal transformation. In Sect.~4 we will show that the condition (\ref{condU}) plays an important role for the equivalence of the Jordan and Einstein frames during inflation.

Let us now perform the conformal transformation
\begin{equation*}
\tilde{g}_{\mu\nu} = 16\pi M_{\mathrm{Pl}}^{-2} U(\phi) g_{\mu\nu},
\end{equation*}
where the metric in the Einstein frame is marked with a tilde, and $M_{\mathrm{Pl}}$ is the Planck mass.

After this transformation, we get a model for a minimally coupled scalar field,
described by the following action
\begin{equation}
S_E =\int d^4x\sqrt{-\tilde{g}}\left[\frac{M_{\mathrm{Pl}}^2}{16\pi}R(\tilde{g}) -
\frac{M_{\mathrm{Pl}}^2}{32\pi U}\left[1+\frac{3{U'}^2}{U}\right]\tilde{g}^{\mu\nu}\phi_{,\mu}\phi_{,\nu}-
V_\mathrm{E}(\phi)\right],
\label{action1}
\end{equation}
where the potential in the Einstein  frame is
\begin{equation}
V_\mathrm{E} = \frac{
M_{\mathrm{Pl}}^4V(\phi)}{256\pi^2U^2(\phi)}.
\label{poten}
\end{equation}

Many inflationary models are based upon the possibility of a slow evolution of some scalar
field $\phi$.  To calculate parameters of inflation that can be tested via observations, we use the slow-roll approximation~\cite{Salopek,Liddle:1994dx} (see also~\cite{Kaiser,Bamba:2014daa}).
During inflation, the slow-roll parameters $\epsilon$, $\eta$ and $\zeta$ should remain to be less than one.
Note that we do not introduce a new scalar field when considering the Einstein frame action, because it is suitable~\cite{EOPV2014} to express slow-roll and inflationary parameters in terms of the initial scalar field $\phi$:
\begin{equation}\label{SLP_phi}
\epsilon = \frac{M_{\mathrm{Pl}}^2{(V_\mathrm{E}')}^2}{16\pi V_\mathrm{E}^2Q}\, ,\qquad
\eta =\frac{M_{\mathrm{Pl}}^2}{8\pi V_\mathrm{E}Q}\left[V_\mathrm{E}''-\frac{V_\mathrm{E}'Q'}{2Q}\right] \, ,
\end{equation}
\begin{equation}
\label{zeta2}
   \zeta^2 =\frac{M_{\mathrm{Pl}}^4 V'_E}{64\pi^2V_\mathrm{E}^2Q^2}\left[V_\mathrm{E}'''-\frac{3V_\mathrm{E}''Q'}{2Q}-
   \frac{V_\mathrm{E}'Q''}{2Q} +\frac{V_\mathrm{E}'(Q')^2}{Q^2}\right],
\end{equation}
where
\begin{equation*}
Q=\frac{M_{\mathrm{Pl}}^2\left(U+3U'^2\right)}{16\pi U^2}.
\end{equation*}
The number of e-foldings in slow-roll inflation is given by the integral~\cite{DeSimone:2008ei}
\begin{equation}
N_e(\phi)=
\frac{8\pi}{M_{\mathrm{Pl}}^2}\int\limits_{\phi_{\mathrm{end}}}^{\phi}
\left|\frac{V_\mathrm{E}}{V'_\mathrm{E}}\right|Q\,d\tilde\phi=\frac{2\sqrt{\pi}}{M_{\mathrm{Pl}}}\int\limits_{\phi_{\mathrm{end}}}^{\phi}
\frac{\sqrt{Q}}{\sqrt{\epsilon}}\,d\tilde\phi\,,
\label{Ne}
\end{equation}
where $\phi_{\mathrm{end}}$ is the value of the field at the end of inflation, defined by the condition $\epsilon=1$ at $\phi=\phi_{\mathrm{end}}$.
The ratio $r$ of squared amplitudes of tensor and scalar perturbations, the scalar spectral index of the primordial curvature fluctuations~$n_\mathrm{s}$, and the
associated running of the spectral index $\alpha_\mathrm{s}$, are
given, to very good approximation, by
\begin{equation}\label{ns}
r = 16 \epsilon\,,\qquad n_\mathrm{s}= 1 - 6 \epsilon + 2
\eta\, , \qquad
\alpha_\mathrm{s} \equiv \frac{d n_\mathrm{s}}{d \ln k}
= 16\epsilon \eta - 24 \epsilon^2 - 2 \zeta^2 .
\end{equation}

\section{Finite gauge  models}

Let us consider a
massless finite or massless asymptotically finite GUT. In flat space-time quantum corrections to the classical potential are either absent or highly suppressed asymptotically.
However,  in curved
space-time in such type of models the coupling parameter $\xi$ corresponding to the
 non-minimal
 scalar-gravitational interaction receives quantum corrections~\cite{9} (for a general review, see~\cite{BOS}).
 In the RG-improvement scheme the corresponding RG-parameter depends on
 the scalar field.
The general structure of the one-loop effective coupling parameter
$\xi (\phi)$ for ``finite'' theories in curved space-time has
been obtained in~\cite{9}. In the one-loop approximation, the RG-equation for the parameter $\xi(\phi)$ is
\begin{equation}
\label{RGequxi}
   \frac{d}{d\vartheta}\xi(\vartheta)=\left(\xi(\vartheta)-\frac{1}{6} \right)Cg^2,
\end{equation}
where $\vartheta(\phi)=\frac{1}{2} \ln \left(\phi^2/\mu^2\right)$, $C$ is a nonzero constant, $\mu$ is a parameter that defines the GUT scale. A parameter $g^2\ll 1$ (it should be clear that $g^2$ is not the square of the determinant of the metric $g_{\mu\nu}$).
Equation~(\ref{RGequxi}) has the following solution:
 \begin{equation}
\label{xi} \xi (\vartheta) = \frac{1}{6} + \left( \xi_0 -
\frac{1}{6} \right) e^{Cg^2\vartheta},
\end{equation}
with an integration constant $\xi_0$.
If $C>0$, then $|\xi (\vartheta)| \rightarrow \infty$
(non-asymptotical conformal invariance) in the UV limit
($\vartheta\rightarrow \infty$). In the models with $C<0$, one gets
$\xi (\vartheta) \rightarrow 1/6$ (asymptotically conformal invariance).

In the model considered, the tree-level functions of the scalar field $V$ and $U$ are~\cite{Elizalde:1993ee}:
\begin{equation}
\label{W0} V^{(0)}= \tilde{a}\lambda \phi^4,\qquad U^{(0)}=b\xi\phi^2,
\end{equation}
where $\tilde{a}$ and $b$ are positive constants and $\xi$ is the
conformal coupling. The corresponding potential in the Einstein frame $V_{\mathrm{E}}$ is a constant and is not suitable for inflation.
To realize the inflationary scenario with a graceful exit from inflation we use the RG-improved potentials\footnote{The main renormalization-group
formulae for models in curved space-time are given in~\cite{Elizalde:1993ee,Elizalde:1994im}. }. The functions $U$ and $V$, obtained in the linear
curvature approximation (see~\cite{Elizalde:1994im,EOPV2014} for details), are given by
\begin{equation}
\label{VUmodi} V=
\tilde{a}\kappa_1 g^2 f^4(\vartheta) \phi^4, \qquad U= b \xi
(\vartheta)f^2(\vartheta)\phi^2,
\end{equation}
where $f(\vartheta)= \exp (-C_1 g^2 \vartheta)$,  $\kappa_1$, and $C_1$ are some constants which depend on the gauge parameter and on the features of the theory, and $\xi (\vartheta)$
is given by (\ref{xi}).  Thus,
\begin{equation*}
f(\vartheta)=\psi^{-C_1g^2},\qquad
\xi(\vartheta)=\frac16+\left(\xi_0-\frac16\right)\psi^{Cg^2},
\end{equation*}
where  $\psi=\phi/\mu$ is a dimensionless field. Observe that, in finite GUTs, the connection
between the parameters $C$ and $C_1$ is not specified.

From (\ref{VUmodi}) we get the potential in the Einstein frame
\begin{equation}\label{Ve to quasi de Sitter}
V_\mathrm{E}(\phi) = \frac{9M_{\mathrm{Pl}}^4\tilde{a}\kappa_1g^2}{64\pi^2b^2\left(1+\left(6\xi_0-1\right)
\psi^{Cg^2}\right)^2}.
\end{equation}
Note that $V_\mathrm{E}$ does not depend on $C_1$, whereas $Q$ depends on it:
\begin{equation*}
Q= \frac{3M_{\mathrm{Pl}}^2\left(b\left[(2+(C-2C_1)g^2)(6\xi_0-1)\psi^{Cg^2}+4-4C_1g^2\right]^2 \psi^{-2C_1g^2}+6+6(6\xi_0-1)\psi^{Cg^2}\right)}{
 16\pi b\psi^{-2C_1g^2}\phi^2\left(1+(6\xi_0-1)\psi^{Cg^2}\right)^2}.
\end{equation*}

In~\cite{EOPV2014} we studied de Sitter solutions in
cosmological models with renormalization-group improved effective
 potentials for some  finite gauge theories. We obtained that there is no de Sitter solutions for $C\neq 0$. In the case $C= 0$ the potential $V_\mathrm{E}$ is a constant. So, this case is not suitable for inflation.

\begin{figure}
    \centering
\includegraphics[height=6.727cm]{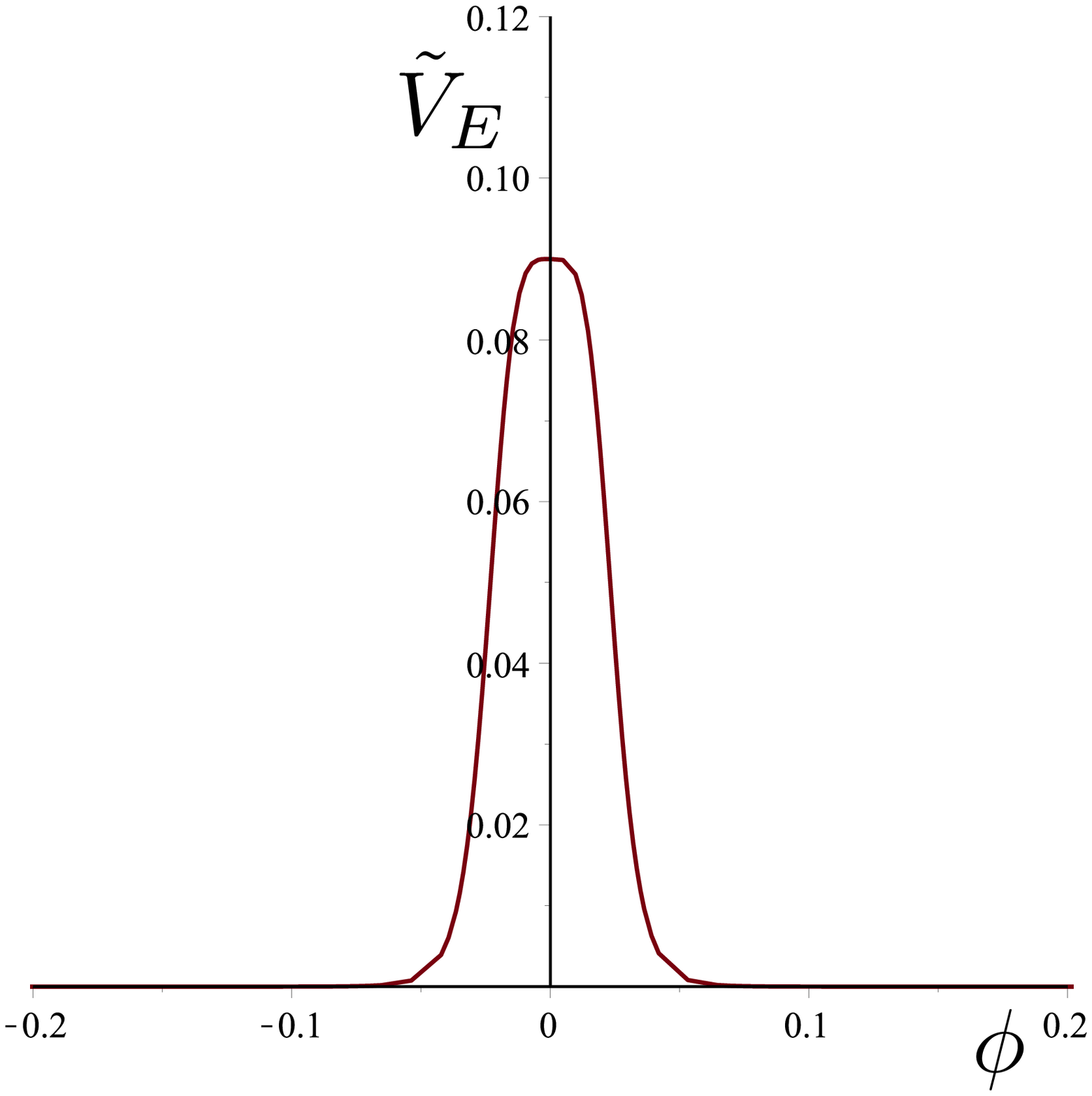} \ \ \  \ \ \ \ \includegraphics[height=6.727cm]{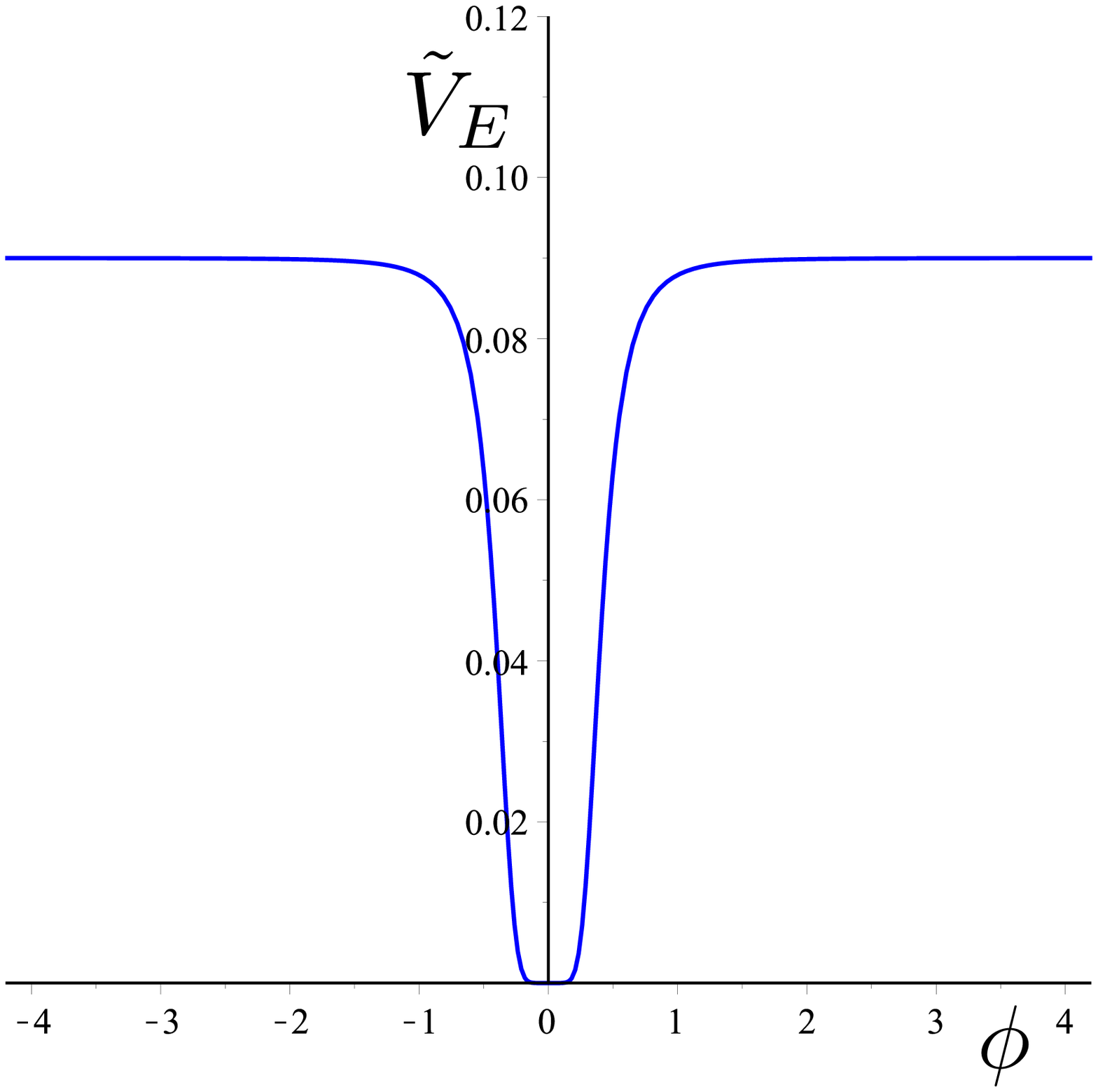}
\caption{The function $\tilde{V}_\mathrm{E}\equiv\frac{64\pi^2}{M_{\mathrm{Pl}}^4}V_\mathrm{E}$ at $C=4/g^2$ (left) and $C=-4/g^2$ (right). Other parameters have the following values: $\xi_0=20$, $\tilde{a}=1$, $b=1$,  $\kappa_1=1$, $g=0.1$, $\mu=0.1$.}
\label{VeFiniteSU2}
\end{figure}

In this paper we consider the possibility to realize an inflationary scenario without an exact de Sitter solution.
We see that $V'_\mathrm{E}$ tends to zero for $\phi \rightarrow\infty$.  When $C<0$ the potential  $V_\mathrm{E}$ tends to a maximal value for $\phi\rightarrow\infty$ (see Fig.~\ref{VeFiniteSU2}) and the model has a quasi-de Sitter solution, in other words,  an approximately constant Hubble parameter for large $\phi$. So, we can expect a slow-roll
behavior of the scalar field.

The functions $U$ and $V$, given by (\ref{VUmodi}), look complicated but for some values of the constants $C$ and $C_1$ they are actually simple and exhibit very interesting properties.  Let us first consider the function $U(\phi)$. The function $U^{(0)}$ corresponds to the induced gravity, in other words, the term proportional to $\phi^2 R$  plays the role of the Hilbert--Einstein term in the action. It is quite interesting that for some values of the parameters the renormalization-group corrections can yield the standard Hilbert--Einstein term. Indeed, if
\begin{equation}
\label{Cg2}
    C_1=\frac{1}{g^2}+\frac{C}{2},
\end{equation}
then  function $U$ acquires the following form
\begin{equation}
\label{UC}
U = \frac{1}{6}b\mu^2\left(6\xi_0-1+\psi^{-Cg^2}\right).
\end{equation}
Thus, the Hilbert--Einstein term arises as a  renormalization-group correction.
If the condition (\ref{Cg2}) is satisfied, then we also obtain
\begin{equation}
\label{VC}
    V=\tilde{a}\kappa_1g^2\mu^4\psi^{-2Cg^2}=\frac{36}{b^2}\tilde{a}g^2\kappa_1\left(U
    -\frac{b\mu^2}{6}(6\xi_0-1)\right)^2.
\end{equation}
This relation between the functions $V$ and $U$ means that the corresponding inflationary model is a cosmological attractor (see the next section).
Dynamics of similar cosmological models, that include the Hilbert--Einstein curvature term and the monomial of the scalar field non-minimally coupled to gravity, have been considered in~\cite{Sami:2012uh}.

Let us analyze possible inflationary scenarios in this case.
As we see from Fig.~\ref{VeFiniteSU2}, the case with a negative $C$ is more suitable for inflation.
Using (\ref{UC}) and (\ref{VC}), we get the inflationary parameters:
\begin{equation}
\label{epsilon}
\epsilon = \frac{4\mu^4b^2(6\xi_0-1)^2{U'}^2}{(b\mu^2(6\xi_0-1)-6U)^2\left(U+3{U'}^2\right)}=\frac{8b n^2(6\xi_0-1)^2}{6n^2b\psi^{4n}+3\psi^{2n+2}+3(6\xi_0-1)\psi^2}.
\end{equation}
\begin{equation*}
\begin{split}
\eta
&=\frac{4nb(6\xi_0-1)}{3\left(2bn^2\psi^{4n}+\psi^{2n+2}
+(6\xi_0-1)\psi^2\right)^2}\left[-4bn^3\psi^{6n}-(n+1)\psi^{4n+2}\right.\\
&\left.{}+4n^3b(6\xi_0-1)\psi^{4n}+
(3n-2)(6\xi_0-1)\psi^{2+2n}+(6\xi_0-1)^2(4n-1)\psi^2\right],
\end{split}
\end{equation*}
where  $n=-Cg^2/2$.

The finite gauge model here considered depends on eight parameters\footnote{Two of them are connected by~\eqref{Cg2}.}, but slow-roll parameters $\epsilon$ and $\eta$ depend only on four parameters, namely $n$, $b$, $\xi_0$, and $\mu$.  The relation between them can be obtained from (\ref{UC}), because the function $U$ should reproduce $M_{\mathrm{Pl}}^2/(16\pi)$ at low energy. Thus,
\begin{equation}
\label{Mpcondition}
    b=\frac{3}{8\pi (6\xi_0-1)\mu^2}M_{\mathrm{Pl}}^2.
\end{equation}
If $\psi^2\gg 1$ during inflation, then slow-roll parameters do not depend on $b$ in first approximation. Therefore, we can fix $b$, using (\ref{Mpcondition}), and  construct suitable inflationary scenarios by choosing the parameters $n$, $\mu$, and $\xi$. We consider a GUT model and thus a reasonable value of $\mu$ is about $10^{-3} M_{\mathrm{Pl}}$. At the same time, it is very  interesting that we can actually get a suitable inflationary scenario with approximately the same values for the inflation parameters, for different values of the model parameters $\xi_0$ and $\mu$. Note that the parameter $\mu$ appears in dimensionless combination $\phi/\mu$ only.

Let us assume that $1/\psi^2$ is a small parameter and $n$ is a natural number. The case $n=1$ is special, because in this case $C_1=0$, the case $n=2$ will be consider below in detail. Let us  consider the  case $n> 2$.  Such as
\begin{equation}
\label{condxi}
\psi^2\gg 1,
\end{equation}
 we obtain from (\ref{epsilon}):
\begin{equation}
\epsilon\approx\frac43\, \left( 6\,{\xi_0}-1 \right)^2{\psi}^{-4n}\,,\quad \eta\approx
{}-\frac{4}{3}(6\xi_0-1)\psi^{-2n}.
\label{appximated epression epsilon}
\end{equation}
In this approximation $\epsilon\mid_{\psi=\psi_{end}}\approx1$
 at
 \begin{equation}
 \label{psiend}
 \psi_{end}\equiv\frac{\phi_{end}}{\mu}\approx\left[\frac{2}{\sqrt{3}}\left( 6\,{\xi_0}-1\right)\right]^{\frac{1}{2n}}.
 \end{equation}

The inflation parameters and the number of e-foldings
 can be approximated as follows
  \begin{equation*}
    n_\mathrm{s}\approx1-\frac83\left( 6\,{\xi_0}-1 \right)\psi_N^{-2n} , \quad
  r \approx {\frac {64}{3}}\, \left( 6\,{\xi_0}-1 \right) ^{2}\psi_N^{-4n},\quad
 \alpha_\mathrm{s} \approx{}-{\frac {32}{9}}\, \left( 6{\xi_0}-1\right)^{2}\psi_N^{-4n},
\end{equation*}
\begin{equation}
N_e \approx\frac{3\left(\psi_{N}^{2n} -\psi_{end}^{2n}\right)}{4\left( 6\,{\xi_0}-1 \right)},
 \end{equation}
  where $\psi_N$ is the value of $\psi$ at which the inflationary parameters are calculated. Using (\ref{psiend}), we get
    \begin{equation*}
         n_\mathrm{s}\approx1-\frac{4}{\sqrt {3}}\,\Upsilon^{-2n}, \quad
      r\approx16\,\Upsilon^{-4n},\quad
    \alpha_\mathrm{s}\approx-\frac83\,\Upsilon^{-4n}\,, \quad N_e\approx\frac{1}{2\sqrt {3}}\left(3{\Upsilon}^{2n}-1 \right),
   \end{equation*}
where $\Upsilon=\psi_{N}/\psi_{end}$.

  For $N_e\simeq 60$, we obtain that $\Upsilon^{2n}\gg1$, hence, an approximated expression for $N_e$ is
   \begin{equation*}
   N_e\approx\frac{\sqrt {3}}{2}\Upsilon^{2n}.
   \end{equation*}
We conclude that in our model, if the condition (\ref{condxi}) is satisfied, then there exist the following relations between the inflation parameters and the e-folding number~$N_e$:
\begin{equation}\label{nsralpha}
  n_\mathrm{s} \approx  1-\frac{2}{N_e},\qquad  r \approx \frac{12}{N_e^2}, \qquad  \alpha_\mathrm{s} \approx {}-\frac{2}{N_e^2}.
\end{equation}

For example, at $N_e=60$, formula (\ref{nsralpha}) gives
\begin{equation}\label{nsralpha60}
    n_\mathrm{s}\approx\frac{29}{30}\approx 0.9667,\qquad r \approx\frac{1}{300}\approx 0.0033333,\qquad
    \alpha_\mathrm{s} \approx {}-\frac{1}{1800}\approx -0.00055556.
\end{equation}

We see that the expressions for $r$ and $n_\mathrm{s}$ in (\ref{nsralpha}) coincide with  (\ref{nsrN}).
Thus, we see that in order to get a inflationary model with inflationary parameters that are in good agreement with
the observational data, it is sufficient to satisfy the conditions $\psi_{N}^{2n}\gg\psi_{end}^{2n}$ and (\ref{condxi}) during inflation.
This result will be explained in the next section with the help of the method the cosmological attractors~\cite{LindeKallosh1}.

Now, we show numerically that the suitable inflationary parameters can be obtained for a wide region of the finite gauge model parameters.  Let us consider the case $n=2$, that corresponds to
\begin{equation}
\label{CC1}
    C_1={}-\frac{1}{g^2},\qquad C={}-\frac{4}{g^2},
\qquad U=\frac{b}{6}\mu^2(6\xi_0-1)+\frac{b}{6\mu^2}\phi^4,\qquad V=\frac{\tilde{a}\kappa_1g^2}{\mu^4}\phi^8.
\end{equation}
Substituting (\ref{CC1}), we obtain
\begin{equation*}
V_\mathrm{E} = \frac{9M_{\mathrm{Pl}}^4\tilde{a}\kappa_1g^2\phi^8}{64\pi^2 b^2\left(\phi^4+(6\xi_0-1)\mu^4\right)^2}\,,\qquad Q = \frac{3M_{\mathrm{Pl}}^2\left[(6\xi_0-1)\mu^6+\mu^2\phi^4+8 b\phi^6\right]}{16\pi b\left(\phi^4+(6\xi_0-1)\mu^4\right)^2},
\end{equation*}
and also
\begin{equation}\label{epsilonpos}
\epsilon = \frac{32b(6\xi_0-1)^2}{3\psi^2\left[3(6\xi_0-1)+\psi^4+8b\psi^6\right]}.
\end{equation}
Thus, we can analytically calculate the value of $\psi$ that corresponds to the end of inflation: $\epsilon=1$.

One can see that, for the values of the parameters given in Table~\ref{SU2InfPara}, the corresponding values of the inflationary parameters $n_\mathrm{s}$, $r$ and $\alpha_\mathrm{s}$ are in good agreement with the observational data (Eqs.~(\ref{nsdata}) and (\ref{rdata})).  Note that the parameters of inflation displayed in Table~\ref{SU2InfPara}
and Table~\ref{InfParamu0001} were obtained through numerical calculation, without
any approximation whatsoever. Parameter $b$ is determined by (\ref{Mpcondition}) and the inflationary parameters depend on the number of e-foldings.

\begin{table}[h]
\begin{center}
\caption{Parameter values for the inflationary scenario at $n=2$.}
\begin{tabular}{|c|c|c|c|c|c|c|c|c|c|c|c|c|}
  \hline
  $\mu$  & $\xi_0$  & $\psi_{end}$            & $ N_e $& $\psi_N $ &   $ n_\mathrm{s}$ & $r $ & ${}-\alpha_\mathrm{s}$ \\
   % &           &  $              & $ $   & $ $       &    $ $   & $ $  & $         $ \\
  \hline
       $ $ &   $ $ & $            $ & $ 50$& $ 9.539$ & $0.9613 $ & $0.0043 $ & $0.00076 $ \\
  $ $    & $  $ & $            $ & $ 55$& $ 9.760$ & $0.9650$ & $0.0036 $ & $0.00063 $ \\
    $0.1M_{\mathrm{Pl}}$  & $20$ & $3.415$     & $ 60$& $ 9.966$ & $0.9675 $ & $0.0030 $ & $0.00053 $ \\
$  $  & $  $ & $             $& $ 65$& $ 10.161$ & $0.9700 $ & $0.0026 $ & $0.00045 $ \\
      \hline
    $ $  & $     $ & $       $ & $50$& $10.083$ & $0.9612 $ & $0.0044 $ & $0.00076 $ \\
  $  $    & $  $ & $            $ & $ 55$& $ 10.317$ & $0.9650 $ & $0.0036 $ & $0.00063 $ \\
   $0.1M_{\mathrm{Pl}}$  & $25$ & $3.523$     & $ 60$& $ 10.536$ & $0.9675 $ & $0.0031 $ & $0.00052 $ \\
$  $   &  & $             $& $ 65$& $ 10.741$ & $0.9700 $ & $0.0026 $ & $0.00045 $ \\
    \hline
    $ $ &   $ $ & $            $                            & $50$ & $67.813$ & $0.9614 $ & $0.0043$  & $0.00075$ \\
  $ $    & $  $ & $       $                              & $ 55$& $69.364$ & $0.9650 $ & $0.0035 $ & $0.00062 $ \\
   $0.01M_{\mathrm{Pl}}$  & $5\cdot10^4$ & $23.961$     & $ 60$& $70.926$ & $0.9678 $ & $0.0030 $ & $0.00052 $ \\
   &  & $             $                                     & $ 65$& $72.199$ & $0.9701 $ & $0.0026 $ & $0.00045 $ \\
          \hline
    $ $ &   $$ & $            $           & $ 50$& $ 4.346$ & $0.9616$ & $0.0042 $ & $0.00075 $ \\
      $ $    & $  $ & $      $ & $ 55$& $ 4.445$ & $0.9650$ & $0.0035 $ & $0.00062 $ \\
      $0.001M_{\mathrm{Pl}}$  & $1$ & $1.550 $     & $ 60$& $ 4.538$ & $0.9678$ & $0.0030 $ & $0.00052 $ \\
 $  $  & $  $ & $  $& $ 65$& $ 4.626$ & $0.9703$ & $0.0025 $ & $0.00045 $ \\
        \hline
    \end{tabular}
\label{SU2InfPara}
\end{center}
\end{table}

From Table~\ref{SU2InfPara} we see that the number of e-foldings $N_e$ essentially influences the values of the inflation parameters, whereas their dependence on $\mu$ and $\xi_0$ is very small. The  values of the inflationary parameters here obtained are close to the values one gets from the approximate formula (\ref{nsralpha60}). This means that a wide domain of parameters $\mu$ and $\lambda$ are actually suitable for inflation. Even then, there are the following restrictions on these parameter. First, we consider $\xi>1/6$, which corresponds to $b>0$ and $U(\phi)>0$ at all $\phi$. To use the approximate formulae for the inflationary parameters we impose the condition (\ref{condxi}), which should be satisfied at the point $\psi=\psi_N$. It is not easy to get the corresponding restriction on the model parameters. At the same time, the stronger condition  $\psi_{end}\gg 1$ gives an explicit restriction  on the parameter $\xi_0$, due to (\ref{psiend}). Note that this condition is actually sufficient. For example, for $\mu=10^{-3}$ the inflation parameters are approximately the same for different values of $\xi_0$ (see Table~\ref{InfParamu0001}). We observe that suitable values for the inflation parameters can be obtained even if $\psi_{end}<1$.

\begin{table}[h]
\begin{center}
\caption{Parameter values for the inflationary scenario at $n=2$, $\mu=0.001M_{\mathrm{Pl}}$ and $ N_e=60 $.}
\begin{tabular}{|c|c|c|c|c|c|c|c|c|c|c|c|c|}
  \hline
 $\mu$  & $\xi_0$  & $\psi_{end}$            & $ N_e $& $\psi_N $ &   $ n_\mathrm{s}$ & $r $ & ${}-\alpha_\mathrm{s}$ \\
   % &           &  $              & $ $   & $ $       &    $ $   & $ $  & $         $ \\
  \hline
     $0.001M_{\mathrm{Pl}}$  & $0.2$ & $0.693$     & $ 60$& $2.030$ & $0.9678$ & $0.0030 $ & $0.00052 $ \\
     \hline
      $0.001M_{\mathrm{Pl}}$  & $0.3$ & $0.980$     & $ 60$& $2.870$ & $0.9678$ & $0.0030 $ & $0.00052$ \\
      \hline
      $0.001M_{\mathrm{Pl}}$  & $1$ & $1.550 $     & $ 60$& $ 4.538$ & $0.9678$ & $0.0030 $ & $0.00052 $ \\
  \hline
     $0.001M_{\mathrm{Pl}}$  & $20$ & $3.424$     & $ 60$& $ 10.024$ & $0.9678$ & $0.0030 $ & $0.00052 $ \\
      \hline
     $0.001M_{\mathrm{Pl}}$  & $10^3$ & $9.123$     & $ 60$& $26.710$ & $0.9678$ & $0.0030 $ & $0.00052 $ \\
     \hline
     $0.001M_{\mathrm{Pl}}$  & $10^5$ & $28.846$     & $ 60$& $ 84.464$ & $0.9679$ & $0.0030 $ & $0.00052 $ \\
     \hline
    \end{tabular}
\label{InfParamu0001}
\end{center}
\end{table}

To get a suitable inflationary scenario it is necessary to obtain inflationary parameters which are compatible with observation data, but this is not sufficient. Indeed, we must examine whether graceful exit from inflation can occur in the  model here considered. It is easy to see that, during inflation, the inflaton moves from large to values of $\phi$ to $\phi=0$, which corresponds to a minimum of the potential in the Einstein frame. Therefore, what we get is slow-roll inflation and subsequent oscillations of the inflaton near a minimum of the potential~$V_{\mathrm{E}}$. We also have shown that, during inflation, our model is close to a well-known inflationary model which has no problems with the exit from inflation. Taking into account that the inflationary scenario was naturally generated by quantum corrections, this is already a quite remarkable result.

\section{Attractor behavior of the considered inflationary model}

There are plenty of inflationary models and it is of interest to find the place of the model here discussed, and to compare it with other known models, what we are going to do next.
In~\cite{LindeKallosh,LindeKallosh1} it has been shown that, if in the Jordan frame $V=c^2(U-U_0)^2$, where $c$ and $U_0$ are some nonzero constants and $V$ is proportional to $\phi$ with some positive power, then one finds a generalization of the Starobinsky potential in the Einstein frame, what has been called the $\alpha-\beta$ attractor model~\cite{LindeKallosh1}.

The idea of cosmological attractors is a very interesting one since it tries to select, among the huge number of inflationary models, a distinguished family of them which is not extremely dependent on the initial conditions and which, in a natural, quite generic setup, may give rise, with high probability, to inflation. It is based on the specific observation that, for many cosmological models with non-minimally coupled scalar fields, the following relation is satisfied:
\begin{equation}
\label{condU}
1\ll\frac{3{U^\prime}^2}{U}.
\end{equation}
In this approximation $Q\thickapprox 3M_{\mathrm{Pl}}^2{U'}^2/(16\pi U^2)$. Also, the action (\ref{action1}) that corresponds to the Einstein frame can be simplified to
\begin{equation*}
%\label{action1a}
\begin{split}
S_E &\simeq\int d^4x\sqrt{-\tilde{g}}\left[\frac{M_{\mathrm{Pl}}^2}{16\pi}R(\tilde{g}) -
\frac{3M_{\mathrm{Pl}}^2{U'}^2}{32\pi U^2}\tilde{g}^{\mu\nu}\phi_{,\mu}\phi_{,\nu}-V_\mathrm{E}(\phi)\right]\\
&\simeq\int d^4x\sqrt{-\tilde{g}}\left[\frac{M_{\mathrm{Pl}}^2}{16\pi}R(\tilde{g}) -
\frac{3M_{\mathrm{Pl}}^2}{32\pi U^2}\tilde{g}^{\mu\nu}U_{,\mu}U_{,\nu}-\tilde{V}_\mathrm{E}(U)\right],
\end{split}
\end{equation*}
where $\tilde{V}_\mathrm{E}(U)=V_\mathrm{E}(\phi(U))$. In other words, one can consider $U$ as a scalar field in the Einstein frame.

Let as assume that the functions $U$ and $V$ are connected by the following relation
\begin{equation}
\label{VUconnect}
V=c^2(U-U_0)^2,\qquad \Rightarrow \qquad \tilde{V}_\mathrm{E} = \frac{M_{\mathrm{Pl}}^4c^2}{256\pi^2}\left(1-\frac{U_0}{U}\right)^2.
\end{equation}
Also, we assume that the function $U$ tends to the constant $U_0$ when $\phi\rightarrow 0$. One can check that indeed the functions $U$ and $V$, given by~(\ref{UC}) and (\ref{VC}), satisfy these conditions. From (\ref{VC}) and (\ref{Mpcondition}) we obtain that for the model considered
\begin{equation}
\label{U0c2}
U_0=\frac{M_\mathrm{Pl}^2}{16\pi},\qquad c^2=256\pi^2\tilde{a}g^2\kappa_1(6\xi_0-1)^2\frac{\mu^4}{M_\mathrm{Pl}^4}.
\end{equation}

To get a scalar field with a standard kinetic term we express $U$ as a function of a new scalar field, $\varphi$:
\begin{equation}
U=U_0e^{\frac{4\sqrt{\pi}}{\sqrt{3}M_{\mathrm{Pl}}}\varphi}.\label{Uchoice}
\end{equation}
Now, the action $S_E$ reads
\begin{equation}
S_E \simeq\int d^4x\sqrt{-\tilde{g}}
\left[\frac{M_{\mathrm{Pl}}^2}{16\pi}R(\tilde{g})-\frac{1}{2}\tilde{g}^{\mu\nu}\varphi_{,\mu}\varphi_{,\nu}-\frac{c^2M_{\mathrm{Pl}}^2}{32\pi}
\left(1-e^{{}-\frac{4\sqrt{\pi}}{\sqrt{3}M_{\mathrm{Pl}}}\varphi}\right)^2\right].
\end{equation}

The  action $S_E$ thus obtained is similar to the corresponding actions in the Einstein frame for the $R^2$ gravity and Higgs-driven inflation models. Indeed, in all cases  the potential has a similar good approximation during inflation, namely
\begin{equation}
V_\mathrm{E}\simeq\tilde{C}\left(1-e^{{}-\frac{4\sqrt{\pi}}{\sqrt{3}M_{\mathrm{Pl}}}\varphi}\right)^2,
\end{equation}
where the constant $\tilde{C}$ is defined in terms of parameters of the model. As a consequence, we come to the conclusion that for values of parameters $C$ and $C_1$ connected by (\ref{Cg2}),
if we can find values for the  other parameters such that that the condition (\ref{condU}) is satisfied, and the value of $\tilde{C}$ is close to the value of the corresponding constant in the $R^2$ gravity and Higgs-driven inflation models, then the inflationary parameters in our model approximately satisfy the relations (\ref{nsrN}). Note that these relations guarantee a good agreement of our inflationary model with Planck data~\cite{Planck2015}. Also, the same conditions ensure a graceful exit from inflation, similar to the exit that occurs in the above-mentioned, well-known inflationary scenarios.

Now, we consider conditions for the strong coupling regime to be satisfied \eqref{condU}.
Substituting the function $U$, given by (\ref{UC}), in \eqref{condU}, we get
\begin{equation}
\frac{2bn^2\psi^{4n-2}}{\psi^{2n}+6\xi_0-1}\gg1.\label{strong coupling realization}
\end{equation}
Using condition  \eqref{Mpcondition}, we obtain
 \begin{equation}
 {\frac{3M_{\mathrm{Pl}}^{2}{\psi}^{4\,n-2}{n}^{2}}{4\pi \, \left( 6\,{\xi_0}
-1 \right) {\mu}^{2} \left( {\psi}^{2n}+6\,{\xi_0}-1 \right) }}\gg 1.
\end{equation}
  For all values of parameters from Table~\ref{InfParamu0001} this inequality is fulfilled and, moreover, the minimal value of the expression on the rhs is higher than $700$. For $\mu=0.1M_{\mathrm{Pl}}$ (see Table~\ref{SU2InfPara}) the value of this expression lays between $4$ and $5$. Note that  the values of the inflationary parameters for $N_e=60$ are similar in both tables.

Let us now consider the family of models found in the preceding section, with arbitrary values for the parameters $C$ and $C_1$. We do not assume that the condition (\ref{Cg2}) is satisfied. Instead, we assume that the functions $U$ and $V$ are connected by Eq.~(\ref{VUconnect}), which
is one of the conditions for the models to belong to the class of the $\alpha-\beta$ cosmological attractors.
We now find the corresponding restrictions on the parameter $C_1$.
Direct substitution of the functions
\begin{equation}
\label{UV}
U=\frac{b}{6}\mu^2\left(1+\left(6\xi_0-1\right)\psi^{Cg^2}\right)\psi^{-2C_1g^2+2}
\quad\mbox{and}
\quad
V=a_1\kappa_1g^2\mu^4\psi^{4(1-C_1g^2)}
\end{equation}
in the condition (\ref{VUconnect}) leads to the equality
\begin{equation*}
\begin{split}
&\left[a_1\kappa_1g^2\mu^4 -\frac{1}{36}\mu^4c^2b^2\right]\psi^{4(1-C_1g^2)}-\frac{(6\xi_0-1)}{18}\mu^4c^2b^2\psi^{4+g^2(C-4C_1)}
\\
-&\frac{(6\xi_0-1)^2}{36}\mu^4c^2b^2\psi^{4+2g^2(C-2C_1)}+\frac{1}{3}\mu^2c^2bU_0\left\{\psi^{2(1-C_1g^2)}+(6\xi_0-1)\psi^{2+g^2(C-2C_1)}\right\}=c^2U_0^2.
\end{split}
\end{equation*}

The r.h.s. of this equation is a nonzero constant, hence, the l.h.s. should include $\psi$ in the zeroth degree. This gives a few possible conditions on the parameters $C$ and $C_1$. One of these condition is (\ref{Cg2}) that allows to construct inflationary models. Let us check other possibilities, namely,
$C_1=1/g^2$ and $C_1=1/g^2+C/4$. The former case corresponds to a constant potential $V$, therefore, condition (\ref{VUconnect}) is satisfied for a constant $U$ only. A straightforward calculation shows that, in the latter case, the condition (\ref{VUconnect}) is satisfied only for $C=0$, then this case coincides with the former one. Thus, in both cases $V_\mathrm{E}$ is a constant and the inflation scenario is not possible.

Summing up, we arrive to the conclusion that condition (\ref{Cg2}) is both necessary and sufficient in order to get the condition (\ref{VUconnect}), which provides a cosmological attractor, known as the $\alpha-\beta$ attractor~\cite{LindeKallosh1}.
 And, as a consequence, we obtain an $\alpha-\beta$ attractor only if the function $U$ includes the Hilbert--Einstein term. But the opposite statement is wrong, because for
\begin{equation*}
    C_1=\frac{1}{g^2},\qquad C=\frac{2}{g^2},
\end{equation*}
we get
\begin{equation*}
U = \frac{1}{6}b\mu^2+(6\xi_0-1)b\phi^2,\qquad V=\tilde{a}\kappa_1 g^2 \mu^4,
\end{equation*}
and the condition (\ref{Cg2}) is not satisfied.

\section{Comparison with $R^2$ gravity}

Condition (\ref{condU})  means, in fact,  that the kinetic term $\frac12g^{\mu\nu}\phi_{,\mu}\phi_{,\nu}$ in action (\ref{action}) is negligibly small during inflation. Thus, inflation in the model here considered  is very close to the inflation model in $f(R)$ gravity that corresponds to the action (\ref{action}) without kinetic term for the scalar field. This fact is very useful in order to analyse the onset of inflation in the  model under consideration.

First of all, for the reasons above, there is actually no difference between the inflationary parameters for the model considered, calculated in the Jordan frame, and the inflationary parameters for the $f(R)$ gravity model, which can be calculated using the Einstein frame. Therefore, if condition (\ref{condU}) is satisfied, then the difference between the inflationary parameters calculated in the Jordan and in the Einstein frame, respectively, is negligibly small.

To construct the inflationary model we make use of the linear curvature approximation, neglecting the induced one-loop pure gravitational term, proportional to $R^2$.  To estimate the importance of this term in the model, we add to action~(\ref{action}) the $\gamma R^2$ term, where $\gamma$ is a constant. Also, we remove the kinetic term that is negligibly small during inflation and get
\begin{equation}
\label{actionm} S_m=\int d^4 x \sqrt{-g}\left[
U(\phi)R-V(\phi)+\gamma R^2\right].
\end{equation}
Using standard formulae~\cite{Faulkner:2006ub}, we obtain the corresponding $f(R)$ gravity:
\begin{equation}
\label{actionR} S_R=\int d^4 x \sqrt{-g}\left[\frac{M_\mathrm{Pl}^2}{16\pi}R+\left(\frac{1}{4c^2}+\gamma\right)R^2\right],
\end{equation}
where the constant $c^2$ is defined by~(\ref{U0c2}). So,  the linear curvature approximation is correct under the condition
$|\gamma|\ll 1/(4c^2)$. For $\mu=10^{-3}$ we obtain
\begin{equation}
\frac{1}{4c^2}=\frac{1}{1024\pi^2\tilde{a}g^2\kappa_1(6\xi_0-1)^2}\times 10^{12}\simeq
\frac{1}{\tilde{a}\kappa_1g^2(6\xi_0-1)^2}\times 10^{8}.
\end{equation}

For Starobinsky's $R^2$ inflation, the coefficient of $R^2$ is defined by the normalization of the amplitude of the
primordial density perturbations, and is of the order of $10^{9}M_\mathrm{Pl}^2$ (see, for example, \cite{Faulkner:2006ub}).
In our model the value of this coefficient gives the condition for their production  $\tilde{a}\kappa_1g^2$.

The coefficient $\gamma$
that corresponds to the induced one-loop pure gravitational
term is much smaller than $10^{8}$. This is a general result for the inflationary model based on quantum field theory.
In Higgs-driven inflation the induced one-loop pure gravitational term, proportional to $R^2$, is negligibly small during the inflationary epoch as well~\cite{BezShapo,DeSimone:2008ei,HiggsRG}. That is why one can neglect the $\gamma R^2$ term and use the linear curvature approach for the construction of inflationary models.

\section{Conclusions}

In this paper we have considered the possibility to construct inflationary models starting from a finite gauge model. The tree-level potential that corresponds to the cosmological constant in the Einstein frame is not suitable for inflation. In our previous work~\cite{EOPV2014}, we checked for the possibility to construct inflationary scenarios with unstable de Sitter solutions and found that this is not possible  for  the finite gauge  models considered.

Here, we have adopted a different strategy and constructed suitable, and quite natural, inflationary scenarios stemming from very fundamental physical principles, and which are perfectly compatible with the most up to date astronomical data. The widely used, standard procedure to get an inflationary model is to add by hand the Hilbert--Einstein term to the action~\cite{HiggsInflation,DeSimone:2008ei,Inagaki:2014wva}. We have here shown that generic RG-corrections allow us to generate this term and to get thus, in a natural way, a model of inflation which is compatible with accurate cosmological observations. For the finite gauge model considered, we have shown explicitly that the inflationary scenario is possible, in spite of the absence of an unstable de Sitter solution. We have also found that, for some reasonable values of the parameters, our finite gauge  model is in good agreement with the most recent  data coming from astronomical observations~\cite{Planck2013,Seljak,Planck2015}.

The inflationary scenario here devised belongs to the class of $\alpha-\beta$ cosmological attractors, according to the convenient classification given in the literature~\cite{LindeKallosh1}. The form of the potential in the Einstein frame is quite close to those of the corresponding potentials for Higgs-driven inflation~\cite{HiggsInflation} and for Starobinsky's $R^2$ inflation~\cite{Starobinsky:1979ty}. This is a remarkable result, taking into account the fact that our model is derived from fundamental physical principles of quantum field theory, which, moreover, lead to the necessary Hilbert--Einstein term in a generic and natural way.

As is known, the inflationary scenarios mentioned above have parameters which are compatible with the astronomical data and very close to one another. This yields even more value to the fact that, at the same time, there is this noticeable difference that distinguishes our model from both $R^2$-inflation and Higgs-driven inflation. The scalar field belongs to the matter sector; moreover, in clear distinction to Higgs-driven inflation, in the model here constructed the Hilbert--Einstein term arises as a result of compulsory quantum corrections at the one-loop approximation and needs not be imposed by hand.

In the paper, we have made explicit comments to the issue of the possible quantum equivalence of the formulations in the Jordan and in the Einstein frames~\cite{Kamenshchik:2014waa,Steinwachs:2011zs}.
We have used RG-improved potential which sums all leading logs beyond the
one-loop approximation. Of course, RG improved effective action includes
also higher-derivative terms, e.g. the $R^2$ term,
which is less relevant for the problem under investigation because it
gives the corrections of next-to-leading order. We are going to
compare the  inflationary models under investigation with the well-known $R^2$
inflationary models with scalar fields~\cite{Gottlober:1993hp} elsewhere.

We have also discussed QFT in curved space-time, in the case where the external gravitational field is a classical one while matter behaves according to QFT. We
have considered Quantum Gravity effects to be less relevant in this approach, since we work below the Planck scale but, generally speaking, the same formulation could be applied to perturbatively renormalizable quantum gravity as, for instance, $R^2$ gravity.  The theory under discussion is multiplicatively renormalizable, as
explicitly demonstrated in the book~\cite{BOS}. We have used a renormalization group formulation for QFT in curved space-time, following~\cite{BOS}, where the
one-loop counter terms are explicitly calculated in dimensional regularization. Owing to the use of dimensional regularization, there is no
dependence on the cut-off and, being multiplicatively renormalizable in the external
gravitational field, the theory under investigation is a closed one (not an
effective theory), since higher-loop corrections repeat the form of the
initial action. Specifically, we work with the effective action calculated in the book~\cite{BOS} and then apply the RG improvement procedure in order to get the sum of all leading logs of the perturbation theory. The calculation of the beta-functions is based on the use of dimensional regularization and, hence, it does not
depend on any explicit cut-off choice. In short, we have worked with the RG improved effective action of the theory under discussion.

Finally, it would be interesting to further compare the finite gauge  inflationary scenario with all these other inflationary models.  To do that, we plan to study the transition from inflation to the later stages of the Universe evolution, starting with reheating~\cite{KLS,ReheatingReview}. As has been shown~\cite{BezrukovR2Higgs}, the reheating temperature can be very different for different models in the same class of cosmological attractors, indeed $T_{reh}=3.1\times10^9 GeV$ for $R^2$ inflation and $T_{reh}\simeq 6\times10^{13} GeV $ \ (with an uncertainty factor of about two) for Higgs-driven inflation. The study of reheating may also give additional constraints on the parameters of the model here considered.  We plan to address the details of the reheating scenario for our class of models in a future publication.

\medskip

\noindent {\bf Acknowledgements}. E.E. and S.D.O. are supported in part by MINECO (Spain), Project FIS2013-44881-P, by the CSIC I-LINK1019 Project, and by the CPAN Consolider Ingenio Project. E.O.P. and S.Yu.V. are supported in part by RFBR according to the research project 14-01-00707 and by the Russian Ministry of Education and Science under grant NSh-3042.2014.2.


\begin{thebibliography}{72}



 \bibitem{Lindebook}
A.D.~Linde, \textit{Particle Physics and Inflationary Cosmology},
Contemporary Concepts in Physics \textbf{5} (1990) 1--362, Harwood Academic, New York,
(arXiv:hep-th/0503203)

\bibitem{19} E.W. Kolb and M.S. Turner, {\it The Early
Universe}, Addison-Wesley, Reading, MA, 1990.

\bibitem{Inflation_review}
 A.D.~Linde,
 \textit{Inflationary Cosmology},
  Lect.\ Notes Phys.\  {\bf 738} (2008) 1
  (arXiv:0705.0164);\\
   J.~Martin, C.~Ringeval, V.~Vennin,
  \textit{Encyclopedia Inflationaris},
  Phys.\ Dark Univ.\  (2014)  (arXiv:1303.3787);\\
  %%CITATION = ARXIV:1303.3787;%%
 A.D.~Linde,
  \textit{Inflationary Cosmology after Planck 2013},
  arXiv:1402.0526
  %%CITATION = ARXIV:1402.0526;%%





  \bibitem{Linde:1981mu}
  A.D.~Linde,
\textit{A New Inflationary Universe Scenario: A Possible Solution
of the Horizon, Flatness, Homogeneity, Isotropy and Primordial
Monopole Problems},
Phys.\ Lett.\ B {\bf 108} (1982) 389;\\
%%CITATION = PHLTA,B108,389;%%
A.D.~Linde,
  \textit{Chaotic Inflation},
Phys.\ Lett.\ B {\bf 129} (1983) 177
%%CITATION = PHLTA,B129,177;%%


\bibitem{Albrecht:1982wi}
A.~Albrecht and P.J.~Steinhardt,
\textit{Cosmology for Grand Unified Theories with Radiatively Induced Symmetry Breaking},
Phys. Rev. Lett.  {\bf 48} (1982)  1220
   %%CITATION = PRLTA,48,1220;%%

\bibitem{Salopek}
D.S. Salopek, J.R. Bond and J.M. Bardeen,
\textit{Designing Density Fluctuation Spectra in Inflation},
Phys. Rev. D \textbf{40} (1989) 1753--1788

\bibitem{inflation2}
J.E. Lidsey, A.R. Liddle, E.W. Kolb, E.J. Copeland, T. Barreiro,
and M. Abney,
\textit{Reconstructing the inflaton potentialan overview},
Rev. Mod. Phys. \textbf{69} (1997) 373--410
(arXiv:astro-ph/9508078);\\
  C.M. Peterson, M. Tegmark,
\textit{Testing Two-Field Inflation},
Phys. Rev. D \textbf{83} (2011) 023522
(arXiv:1005.4056);\\
Shi Pi, M. Sasaki,
\textit{Curvature perturbation spectrum in two-field inflation with a
turning trajectory},
J. Cosmol. Astropart. Phys. \textbf{1210} (2012) 051
(arXiv:1205.0161);\\
P.~Creminelli, D.L.~Nacir, M.~Simonovic, G.~Trevisan and
M.~Zaldarriaga,
  \textit{$\phi^2$ Inflation at its Endpoint}, Phys. Rev. D \textbf{90} (2014) 083513 (arXiv:1405.6264)

\bibitem{NO2011}
S.~Nojiri and S.D.~Odintsov,
 \textit{Unified cosmic history in modified gravity: from ${\cal F}(R)$
theory to Lorentz non-invariant models},
Phys. Rept. \textbf{505} (2011) 59--144 (arXiv:1011.0544);\\
%%CITATION = ARXIV:1011.0544;%%
S.~Nojiri and S.D.~Odintsov,
  \textit{Inflation without self-reproduction in $F(R)$ gravity},
  Astrophys.\ Space Sci.\  {\bf 357} (2015) 1,  39
  (arXiv:1412.2518).
  %%CITATION = ARXIV:1412.2518;%%

\bibitem{Faulkner:2006ub}
  T.~Faulkner, M.~Tegmark, E.F.~Bunn and Y.~Mao,
  \textit{Constraining f(R) Gravity as a Scalar Tensor Theory},
  Phys.\ Rev.\ D {\bf 76} (2007) 063505
  (astro-ph/0612569)
  %%CITATION = doi:10.1103/PhysRevD.76.063505;%%

\bibitem{Sebastiani:2015kfa}
 M.~Rinaldi, G.~Cognola, L.~Vanzo and S.~Zerbini, \textit{Inflation in scale-invariant theories of gravity},
Phys.\ Rev.\ D {\bf 91} (2015)  123527
(arXiv:1410.0631);\\
%%CITATION = ARXIV:1410.0631;%%
  L.~Sebastiani and R.~Myrzakulov,
  \textit{F(R) gravity and inflation}, Int. J. Geom. Meth. Mod. Phys. \textbf{12} (2015) 1530003
  (arXiv:1506.05330)
  %%CITATION = ARXIV:1506.05330;%%

\bibitem{Starobinsky:1979ty}
 A.A.~Starobinsky,
 \textit{Relict Gravitation Radiation Spectrum and Initial State
of the Universe} (In Russian),
JETP Lett.\ {\bf 30} (1979) 682 [Pisma Zh.\ Eksp.\
Teor.\ Fiz.\ {\bf 30} (1979) 719--723];\\
  %%CITATION = JTPLA,30,682;%%
  A.A.~Starobinsky,
 \textit{A New Type of Isotropic Cosmological Models Without
Singularity},
    Phys.\ Lett. B {\bf 91} (1980)  99--102;\\
A.A.~Starobinsky, Lect. Notes in Phys. \textbf{246}
(1986) 107;\\
  %%CITATION = PHLTA,B91,99;%%
A. Vilenkin,   \textit{Classical and Quantum Cosmology of the Starobinsky Inflationary Model},  Phys. Rev. D {\bf 32} (1985)  2511

\bibitem{Mukhanov:1981xt}
  V.F.~Mukhanov and G.V.~Chibisov,
 \textit{Quantum Fluctuation and Nonsingular Universe} (In
Russian),
JETP Lett.\ {\bf 33} (1981) 532--535, [Pisma Zh.\
Eksp.\ Teor.\ Fiz.\ {\bf 33} (1981) 549--553].
  %%CITATION = JTPLA,33,532;%%


\bibitem{nonmin-quant}
A.O. Barvinsky  and A.Yu. Kamenshchik,
 \textit{Quantum scale of inflation and particle physics of the early universe},
Phys. Lett.  B {\bf 332} (1994) 270  (arXiv:gr-qc/9404062)

\bibitem{Cervantes-Cota1995}
   J.L.~Cervantes-Cota and H.~Dehnen,
 \textit{Induced gravity inflation in the standard model of
particle physics},
Nucl.\ Phys. B {\bf 442} (1995) 391
(arXiv:astro-ph/9505069)
  %%CITATION = ASTRO-PH/9505069;%%



\bibitem{Lyth:1998xn}
  D.H.~Lyth and A.~Riotto,
\textit{Particle physics models of inflation and the cosmological
density perturbation},
  Phys.\ Rept.\  {\bf 314} (1999) 1--146 (arXiv:hep-ph/9807278);\\
  %%CITATION = HEP-PH/9807278;%%
K. Kannike, A. Racioppi, and M. Raidal,
\textit{Embedding inflation into the Standard Model -
more evidence for classical scale invariance},
J. High Energy Phys. \textbf{1406} (2014) 154, (arXiv:1405.3987).

\bibitem{BezShapo}
F.L.~Bezrukov and M.~Shaposhnikov,
\textit{The Standard Model Higgs boson as the inflaton},
Phys. Lett. B \textbf{659} (2008) 703 (arXiv:0710.3755)

\bibitem{HiggsInflation}
A.O.~Barvinsky, A.Y.~Kamenshchik, and A.A.~Starobinsky,
\textit{Inflation scenario via the Standard Model Higgs boson
and LHC}, J. Cosmol. Astropart. Phys. {\bf 0811} (2008) 021
(arXiv:0809.2104);\\
 F.~Bezrukov, D.~Gorbunov and M.~Shaposhnikov,
   \textit{On initial conditions for the Hot Big Bang},
J. Cosmol. Astropart. Phys. {\bf 0906} (2009) 029,
(arXiv:0812.3622);\\
  %%CITATION = ARXIV:0812.3622;%%
F.L.~Bezrukov, A.~Magnin, and M.~Shaposhnikov,
\textit{Standard Model Higgs boson mass from inflation},
Phys. Lett. B {\bf 675} (2009) 88 (arXiv:0812.4950);\\
 A.O.~Barvinsky, A.Y.~Kamenshchik, C.~Kiefer, A.A.~Starobinsky,
and C.F.~Steinwachs,
 \textit{Asymptotic freedom in inflationary cosmology with a
nonminimally coupled Higgs field},
J. Cosmol. Astropart. Phys. \textbf{0912} (2009) 003
(arXiv:0904.1698);\\
J.~Garcia-Bellido, D.G.~Figueroa, and J.~Rubio,
\textit{Preheating in the Standard Model with the Higgs-Inflaton
coupled to gravity},
Phys. Rev. D \textbf{79} (2009) 063531
(arXiv:0812.4624);\\
 F.~Bezrukov and M.~Shaposhnikov,
  \textit{Standard Model Higgs boson mass from inflation: Two loop analysis},
  J. High Energy Phys. {\bf 0907} (2009) 089
 (arXiv:0904.1537);\\
F.L.~Bezrukov, A.~Magnin, M.~Shaposhnikov and S.~Sibiryakov,
\textit{Higgs inflation: consistency and generalisations}
J. High Energy Phys. {\bf 1101} (2011) 016
(arXiv:1008.5157);\\
F.~Bezrukov,
\textit{The Higgs field as an inflaton},
Class. Quant. Grav. {\bf 30} (2013) 214001
(arXiv:1307.0708);\\
K.~Allison,
  \textit{Higgs xi-inflation for the 125--126 GeV Higgs: a two-loop analysis},
  J. High Energy Phys. {\bf 1402} (2014) 040 (arXiv:1306.6931);\\
Y.~Hamada, H.~Kawai, K.-y.~Oda and S.C.~Park,
  \textit{Higgs Inflation is Still Alive after the Results from BICEP2},
  Phys. Rev. Lett. \textbf{112} (2014) 241301 (arXiv:1403.5043);\\
   F.~Bezrukov and M.~Shaposhnikov,
  \textit{Higgs inflation at the critical point},
  Phys.\ Lett.\ B {\bf 734} (2014) 249
  (arXiv:1403.6078);\\
  %%CITATION = ARXIV:1403.6078;%%
Hong-Jian He and Zhong-Zhi Xianyu, \textit{Extending Higgs Inflation with TeV Scale New Physics},
  J. Cosmol. Astropart. Phys. {\bf 1410} (2014) 019
  (arXiv:1405.7331)


\bibitem{DeSimone:2008ei}
  A.~De Simone, M.P.~Hertzberg and F.~Wilczek,
  \textit{Running Inflation in the Standard Model},
  Phys.\ Lett.\ B {\bf 678} (2009) 1
  (arXiv:0812.4946)
  %%CITATION = ARXIV:0812.4946;%%


\bibitem{HiggsRG}
A.O.~Barvinsky, A.Yu.~Kamenshchik, C.~Kiefer, A.A.~Starobinsky,
and C.F.~Steinwachs,
\textit{Higgs boson, renormalization group, and cosmology},
Eur. Phys. J. C \textbf{72} (2012) 2219
(arXiv:0910.1041)

\bibitem{Lerner}
R.N. Lerner and J. McDonald,
\textit{Higgs Inflation and Naturalness},
J. Cosmol. Astropart. Phys.  \textbf{1004} (2010) 015,  (arXiv:0912.5463);\\
J.~Ren, Z.-Z.~Xianyu, H.-J.~He,
  \textit{Higgs Gravitational Interaction, Weak Boson Scattering, and Higgs Inflation in Jordan and Einstein Frames},
J. Cosmol. Astropart. Phys. {\bf 1406} (2014) 032 (arXiv:1404.4627)

\bibitem{SUSEinflation}
B.A.~Ovrut and P.J.~Steinhardt,
\textit{Supersymmetric Inflation, Baryon Asymmetry and the Gravitino
Problem},
  Phys.\ Lett.\ B {\bf 147} (1984) 263;\\
  %%CITATION = PHLTA,B147,263;%%
G.R.~Dvali,
\textit{Natural inflation in SUSY and gauge mediated curvature of the
flat directions},  Phys.\ Lett.\ B {\bf 387} (1996) 471
  (arXiv:hep-ph/9605445);\\
   L.~Alvarez-Gaume, C.~Gomez and R.~Jimenez,
  \textit{A Minimal Inflation Scenario},
  J. Cosmol. Astropart. Phys. {\bf 1103} (2011) 027
  (arXiv:1101.4948);\\
  %%CITATION = ARXIV:1101.4948;%%
   C.~Pallis,
\textit{Induced-Gravity Inflation in no-Scale Supergravity and
Beyond}, J. Cosmol. Astropart. Phys.  1408 (2014) 057 (arXiv:1403.5486)
  %%CITATION = ARXIV:1403.5486;%%

\bibitem{Carrasco:2015pla}
  J.~J.~M.~Carrasco, R.~Kallosh and A.~Linde,
  \textit{$\alpha $-Attractors: Planck, LHC and Dark Energy}, J. High Energy Phys. 1510 (2015) 147 (arXiv:1506.01708)
  %%CITATION = ARXIV:1506.01708;%%

\bibitem{Pallis:2013yda}
  C.~Pallis,
  \textit{Linking Starobinsky-Type Inflation in no-Scale Supergravity to MSSM},
  J. Cosmol. Astropart. Phys. {\bf 1404} (2014) 024
  (arXiv:1312.3623);\\
  L.~E.~Ibanez, F.~Marchesano and I.~Valenzuela,
  \textit{Higgs-otic Inflation and String Theory},
  J. High Energy Phys. {\bf 1501} (2015) 128
  (arXiv:1411.5380);\\
   S.~Bielleman, L.~E.~Ibanez, F.~G.~Pedro and I.~Valenzuela,
  \textit{Multifield Dynamics in Higgs-Otic Inflation},
  arXiv:1505.00221.
  %%CITATION = ARXIV:1505.00221;%%


\bibitem{GUT_Inflation}
 J.L.~Cervantes-Cota and H.~Dehnen,
 \textit{Induced gravity inflation in the SU(5) GUT},
Phys.\ Rev. D {\bf 51} (1995) 395
(arXiv:astro-ph/9412032);\\
  %%CITATION = ASTRO-PH/9412032;%%
  S.~Dimopoulos, G.R.~Dvali and R.~Rattazzi,
  \textit{Dynamical inflation and unification scale on quantum moduli spaces},
  Phys.\ Lett.\ B {\bf 410} (1997) 119
  (arXiv:hep-ph/9705348);\\
  %%CITATION = HEP-PH/9705348;%%
 M.B.~Einhorn and D.R.T.~Jones,
  \textit{GUT Scalar Potentials for Higgs Inflation},
  J. Cosmol. Astropart. Phys. {\bf 1211} (2012) 049
  (arXiv:1207.1710);\\
   F.~Brummer, V.~Domcke and V.~Sanz,
  \textit{GUT-scale inflation with sizeable tensor modes}, J. Cosmol. Astropart. Phys. {\bf  1408} (2014) 066 (arXiv:1405.4868)



\bibitem{BOS} I.L. Buchbinder, S.D. Odintsov and I.L. Shapiro,
 \textit{Effective Action in Quantum Gravity}, IOP Publishing, Bristol
and Philadelphia, 1992.

\bibitem{BO1985}
 I.L. Buchbinder, S.D. Odintsov,
\textit{Effective potential and phase transitions induced by curvature
in gauge theories in curved spacetime},
  Class. Quant. Grav. \textbf{2} (1985) 721--731

\bibitem{Elizalde:1993ee}
  E.~Elizalde and S.D.~Odintsov,
\textit{Renormalization group improved effective potential for gauge
theories in curved space-time},
Phys.\ Lett.\ B {\bf 303} (1993) 240  (arXiv:hep-th/9302074);\\
  E. Elizalde and S.D. Odintsov,
\textit{Renormalization group improved effective Lagrangian for
interacting theories in curved space-time},
 Phys. Lett. B \textbf{321} (1994) 199--204 (arXiv:hep-th/9311087)


\bibitem{Elizalde:1994im}
  E.~Elizalde and S.D.~Odintsov,
\textit{Renormalization group improved effective potential for finite
grand unified theories in curved space-time},
  Phys.\ Lett.\ B {\bf 333} (1994) 331
  (arXiv:hep-th/9403132)
  %%CITATION = HEP-TH/9403132;%%


\bibitem{Planck2013}
  P.A.R.~Ade {\it et al.} [Planck Collaboration],
  \textit{Planck 2013 results. XXII. Constraints on inflation},
  Astron.\ Astrophys.\  {\bf 571} (2014) A22
  (arXiv:1303.5082);\\
  %%CITATION = ARXIV:1303.5082;%%\\
P.A.R. Ade, {\it et. al.} [Planck Collaboration],
  \textit{Planck 2013 Results. XXIV. Constraints on primordial non-Gaussianity},
  Astron.\ Astrophys.\  {\bf 571} (2014) A24
  (arXiv:1303.5084);\\
  %%CITATION = ARXIV:1303.5084;%%
%\bibitem{PlanckBicep}
  P.A.R.~Ade {\it et al.}  [BICEP2 and Planck Collaborations],
  \textit{Joint Analysis of BICEP2/$Keck Array$ and $Planck$ Data},
  Phys.\ Rev.\ Lett.\  {\bf 114}, 101301 (2015)
  (arXiv:1502.00612).
  %%CITATION = ARXIV:1502.00612;%%

\bibitem{Planck2015}
P.A.R. Ade, {\it et. al.} [Planck Collaboration],
  \textit{Planck 2015 results. XX. Constraints on inflation},
   arXiv:1502.02114


\bibitem{Coleman} S. Coleman and E. Weinberg, \textit{Radiative Corrections as the Origin of Spontaneous Symmetry Breaking},
 Phys. Rev. D {\bf 7} (1973) 1888.

\bibitem{Stelle} P.S. Howe, K.S. Stelle and P.K. Townsend, \textit{The Relaxed Hypermultiplet: An Unconstrained N=2 Superfield Theory}, Nucl. Phys. B \textbf{214} (1983) 519;\\
P.S. Howe, K.S. Stelle and P.K. Townsend, \textit{Miraculous Ultraviolet Cancellations in Supersymmetry Made Manifest}, Nucl. Phys.  B \textbf{236} (1984)
125

\bibitem{Ermushev8} A.V. Ermushev, D.I. Kazakov and O.V.
Tarasov, \textit{Finite $N=1$ Supersymmetric Grand Unified Theories},
Nucl. Phys. B {\bf 281} (1987) 72;\\
D. Kapetanakis, M.
Mondrag\'on
and G. Zoupanos, \textit{Finite unified models}, Z. Phys. C {\bf 60} (1993) 181;\\
  S.~Heinemeyer, M.~Mondragon and G.~Zoupanos,
  \textit{Finite Theories Before and After the Discovery of a Higgs Boson at the LHC},
  Fortsch.\ Phys.\  {\bf 61} (2013) 11,  969
  (arXiv:1305.5073).
  %%CITATION = ARXIV:1305.5073;%%


\bibitem{WMAP}
 D.N.~Spergel {\it et al.}  [WMAP Collaboration],
 \textit{First Year Wilkinson Microwave Anisotropy Probe (WMAP) Observations:
 Determination of Cosmological Parameters},
 Astrophys.\ J.\ Suppl. {\bf 148} (2003)  175--194
 (arXiv:astro-ph/0302209);  \\
 D.N.~Spergel {\it et al.}  [WMAP Collaboration],
 \textit{Wilkinson Microwave Anisotropy Probe (WMAP) three year results:
 Implications for cosmology},
 Astrophys.\ J.\ Suppl. {\bf 170} (2007)
 377 (arXiv:astro-ph/0603449); \\
 %%CITATION = APJSA,170,377;%%
E.~Komatsu {\it et al.}  [WMAP Collaboration],
 \textit{Seven-Year Wilkinson Microwave Anisotropy Probe (WMAP) Observations:
  Cosmological Interpretation},
  Astrophys.\ J.\ Suppl.  {\bf 192} (2011) 18
  (arXiv:1001.4538);\\
  %%CITATION = APJSA,192,18;%%
  G.~Hinshaw {\it et al.}  [WMAP Collaboration],
\textit{Nine-Year Wilkinson Microwave Anisotropy Probe (WMAP)
Observations: Cosmological Parameter Results},
Astrophys.\ J.\
Suppl.\ {\bf 208} (2013) 19 (arXiv:1212.5226)



\bibitem{Seljak}
  M.J.~Mortonson and U.~Seljak,
  \textit{A joint analysis of Planck and BICEP2 B modes including dust polarization uncertainty},
  J. Cosmol. Astropart. Phys. {\bf 1410} (2014) 10,  035
  (arXiv:1405.5857).
  %%CITATION = ARXIV:1405.5857;%%

\bibitem{GB2013}
F.~Bezrukov, D.~Gorbunov,
\textit{Light inflaton after LHC8 and WMAP9 results},
J. High Energy Phys.
\textbf{1307} (2013) 140  (arXiv:1303.4395)

\bibitem{KL2013}
R.~Kallosh, A.~Linde,
\textit{Superconformal generalization of the chaotic inflation
model $\frac{\lambda}{4} \phi^{4} - \frac{\xi}{2} \phi^{2}R$},
J. Cosmol. Astropart. Phys. {\bf 1306} (2013) 027
(arXiv:1306.3211)
 %%CITATION = ARXIV:1306.3211;%%

\bibitem{Chernikov}
  N.A. Chernikov,  E.A. Tagirov,
\textit{Quantum theory of scalar fields in de Sitter space-time},
  Annales Poincare Phys.\ Theor.\ A  \textbf{9} (1968) 109;\\
    %%CITATION = AHPAA,A9,109;%%
  E.A. Tagirov,
  \textit{Consequences of field quantization in de Sitter type cosmological models},
  Annals Phys. \textbf{76} (1973) 561
  %%CITATION = APNYA,76,561;%%


\bibitem{nonmin-infl}
B.L. Spokoiny,
\textit{Inflation And Generation Of Perturbations In Broken Symmetric Theory Of Gravity},
Phys. Lett. B \textbf{147} (1984) 39--43;\\
T. Futamase and K.-i.  Maeda,
\textit{Chaotic Inflationary Scenario In Models Having Nonminimal Coupling With Curvature},
Phys. Rev. D \textbf{39} (1989) 399--404;\\
R. Fakir and W.G. Unruh,
\textit{Improvement on cosmological chaotic inflation through nonminimal coupling},
Phys. Rev. D \textbf{41} (1990) 1783--1791;\\
M.V.~Libanov, V.A.~Rubakov and P.G.~Tinyakov,
 \textit{Cosmology with nonminimal scalar field: Graceful entrance into inflation},
Phys. Lett. B {\bf 442} (1998) 63 (arXiv:hep-ph/9807553)



\bibitem{OdintsovNOnmin}
T. Muta, S.D. Odintsov,
 \textit{Model dependence of the nonminimal scalar graviton effective
coupling constant in curved space-time},
Mod. Phys. Lett. A \textbf{6} (1991) 3641--3646;\\
S. Mukaigawa, T. Muta, S.D. Odintsov,
\textit{Finite grand unified theories and inflation},
Int. J. Mod. Phys. A \textbf{13} (1998) 2739--2746 (arXiv:hep-ph/9709299)

\bibitem{KKhT} A.Yu. Kamenshchik, I.M. Khalatnikov, and A.V.
Toporensky,
\textit{Complex inflaton field in quantum cosmology},
Int. J. Mod. Phys. D \textbf{6} (1997) 649--672
(arXiv:gr-qc/9801039)


\bibitem{Cerioni}
A. Cerioni, F. Finelli, A. Tronconi and G. Venturi,
\textit{Inflation and Reheating in Induced Gravity},
Phys. Lett. B \textbf{681} (2009) 383--386
(arXiv:0906.1902); \\
A. Cerioni, F. Finelli, A. Tronconi and G. Venturi,
 \textit{Inflation and Reheating in Spontaneously Generated
Gravity},
Phys. Rev. D \textbf{81} (2010) 123505 (arXiv:1005.0935);
\\
A. Tronconi and G. Venturi,
\textit{Quantum Back-Reaction in Scale Invariant Induced Gravity
Inflation},
Phys. Rev. D \textbf{84} (2011) 063517 (arXiv:1011.39580)



\bibitem{Cooper:1982du}
  F.~Cooper and G.~Venturi,
  \textit{Cosmology and broken scale invariance},
 Phys. Rev.  D \textbf{24} (1981) 3338


\bibitem{Kaiser}
D.I.~Kaiser,
 \textit{Induced-gravity Inflation and the Density Perturbation
Spectrum},
Phys. Lett. B \textbf{340} (1994) 23--28
(arXiv:astro-ph/9405029);\\
%%CITATION = ASTRO-PH/9405029;%%
D.I.~Kaiser,
\textit{Primordial spectral indices from generalized Einstein
theories},
Phys. Rev. D {\bf 52} (1995) 4295
(arXiv:astro-ph/9408044)
%%CITATION = ASTRO-PH/9408044;%%





\bibitem{EOPV2014}
E. Elizalde, S.D. Odintsov, E.O.~Pozdeeva, and S.Yu.~Vernov,
\textit{Renormalization-group inflationary scalar electrodynamics
and $SU(5)$ scenarios confronted with Planck2013 and BICEP2 results},
Phys. Rev. D \textbf{90} (2014) 084001
(arXiv:1408.1285)


\bibitem{Inagaki:2014wva}
  T.~Inagaki, R.~Nakanishi and S.D.~Odintsov,
  \textit{Inflationary Parameters in Renormalization Group Improved $\phi^4$ Theory},
  Astrophys. Space Sci. \textbf{354} (2014) 2, 2108 (arXiv:1408.1270);\\
  %%CITATION = ARXIV:1408.1270;%%
T.~Inagaki, R.~Nakanishi and S.~D.~Odintsov,
  \textit{Non-Minimal Two-Loop Inflation},
  Phys.\ Lett.\ B {\bf 745} (2015) 105
  (arXiv:1502.06301)
  %%CITATION = ARXIV:1502.06301;%%




 \bibitem{LindeKallosh}
M. Galante, R. Kallosh, A. Linde and D. Roest,
\textit{A universal attractor for inflation at strong coupling},
Phys. Rev. Lett. \textbf{112} (2014) 011303 (arXiv:1310.3950)

\bibitem{Kallosh:2014rha}
  R.~Kallosh,
  \textit{More on Universal Superconformal Attractors},
  Phys.\ Rev.\ D {\bf 89} (2014)  087703
  (arXiv:1402.3286)
  %%CITATION = ARXIV:1402.3286;%%

\bibitem{LindeKallosh1}
M. Galante, R. Kallosh, A. Linde and D. Roest,
\textit{The Unity of Cosmological Attractors},
Phys. Rev. Lett. \textbf{114} (2015)  141302
(arXiv:1412.3797)

\bibitem{LindeKallosh2}  R.~Kallosh, A.~Linde and D.~Roest,
  \textit{The double attractor behavior of induced inflation},
  J. High Energy Phys. {\bf 1409} (2014) 062
  (arXiv:1407.4471)





\bibitem{rgi}
R.P.~Woodard, \textit{Cosmology is not a Renormalization Group Flow},
  Phys.\ Rev.\ Lett.\  {\bf 101} (2008) 081301
  (arXiv:0805.3089);\\
 H.M.~Lee, \textit{Running inflation with unitary Higgs},
  Phys.\ Lett.\ B {\bf 722} (2013) 198
  (arXiv:1301.1787); \\
 N.~Okada and Q.~Shafi, \textit{Observable Gravity Waves From $U(1)_B-L$ {B-L} Higgs and Coleman-Weinberg
      Inflation}, arXiv:1311.0921 [hep-ph];\\
G.~Barenboim, E.J.~Chun and H.M.~Lee, \textit{Coleman--Weinberg Inflation in light of Planck},
  Phys.\ Lett.\ B {\bf 730} (2014) 81;\\
  I.~Oda and T.~Tomoyose, \textit{Quadratic Chaotic Inflation from Higgs Inflation},
  Adv.\ Stud.\ Theor.\ Phys.\  {\bf 8} (2014) 551
  (arXiv:1404.1538);\\
  Y.~Hamada, H.~Kawai and K.y.~Oda, \textit{Predictions on mass of Higgs portal scalar dark matter from Higgs
      inflation and flat potential},
  J. High Energy Phys. {\bf 1407} (2014) 026;\\
     Y.~Hamada, H.~Kawai, K.y.~Oda and S.C.~Park, \textit{Higgs inflation from Standard Model criticality},
  Phys.\ Rev.\ D {\bf 91} (2015) 053008
  (arXiv:1408.4864); \\
    M.~Herranen, T.~Markkanen, S.~Nurmi and A.~Rajantie, \textit{Spacetime curvature and the Higgs stability during inflation},
  Phys.\ Rev.\ Lett.\  {\bf 113} (2014)   211102
  (arXiv:1407.3141); \\
  M.~Herranen, A.~Osland and A.~Tranberg, \textit{Quantum corrections to inflaton dynamics, the semi-classical approach and
      the semi-classical limit,} Phys.Rev. D \textbf{92} (2015) 083530 (arXiv:1503.07661);\\
K. Kannike, A. Racioppi, and M. Raidal, \textit{Linear inflation from quartic potential}, J. High Energy Phys. {\bf  1601} (2016) 035 (arXiv:1509.05423)

\bibitem{Inagaki:2015eza}
  T.~Inagaki, S.D.~Odintsov and H.~Sakamoto,
  \textit{Gauged Nambu-Jona-Lasinio inflation}, Astrophys. Space Sci. \textbf{360} (2015) 2, 67
  (arXiv:1509.03738)
  %%CITATION = ARXIV:1509.03738;%%


\bibitem{BezrukovR2Higgs}
 F.L.~Bezrukov and D.S.~Gorbunov,
  \textit{Distinguishing between $R^2$-inflation and Higgs-inflation},
  Phys.\ Lett.\ B {\bf 713} (2012) 365
  (arXiv:1111.4397)
  %%CITATION = ARXIV:1111.4397;%%

\bibitem{Muh}  V.~Mukhanov,
  \textit{Quantum Cosmological Perturbations: Predictions and Observations}
  Eur.\ Phys.\ J.\ C {\bf 73} (2013) 2486   (arXiv:1303.3925)

\bibitem{Roest:2013fha}
  D.~Roest,
  \textit{Universality classes of inflation},
  J. Cosmol. Astropart. Phys.  {\bf 1401} (2014) 007
  (arXiv:1309.1285)
  %%CITATION = doi:10.1088/1475-7516/2014/01/007;%%

\bibitem{Binetruy:2014zya}
P.~Binetruy, E.~Kiritsis, J.~Mabillard, M.~Pieroni and C.~Rosset,
\textit{Universality classes for models of inflation},
J. Cosmol. Astropart. Phys.  {\bf 1504} (2015) 033 (arXiv:1407.0820)

\bibitem{Ventury2015}
M.~Rinaldi, L.~Vanzo, S.~Zerbini and G.~Venturi, \textit{Inflationary quasi-scale invariant attractors},
arXiv:1505.03386 [hep-th].
%%CITATION = ARXIV:1505.03386;%%

\bibitem{Pieroni2015}
 M.~Pieroni,
  \textit{$\beta$-function formalism for inflationary models with a non minimal coupling with gravity},
  arXiv:1510.03691 [hep-ph].
  %%CITATION = ARXIV:1510.03691;%%

\bibitem{KTV2011}
A.Yu. Kamenshchik, A. Tronconi, and G. Venturi,
\textit{Reconstruction of scalar potentials in induced gravity
and cosmology},
Phys. Lett. B \textbf{702} (2011) 191--196
(arXiv:1104.2125);\\
A.Yu.~Kamenshchik, A.~Tronconi, G.~Venturi, and S.Yu.~Vernov,
\textit{Reconstruction of Scalar Potentials in Modified Gravity
Models},
Phys. Rev. D \textbf{87} (2013) 063503 (arXiv:1211.6272)

\bibitem{frames}
V. Faraoni and Sh. Nadeau,
\textit{(Pseudo)issue of the conformal frame revisited},
Phys. Rev. D \textbf{75} (2007) 023501 (arXiv:gr-qc/0612075);\\
S. Capozziello, S. Nojiri, S.D. Odintsov and A. Troisi,
\textit{Cosmological viability of $f(R)$-gravity as an ideal fluid and
its compatibility with a matter dominated phase,}
Phys.\ Lett. B \textbf{639} (2006) 135 (arXiv:astro-ph/0604431);\\
M.P. Hertzberg,
\textit{On Inflation with Non-minimal Coupling},
J. High Energy Phys. 1011 (2010) 023  (arXiv:1002.2995);\\
D.I. Kaiser,
\textit{Conformal Transformations with Multiple Scalar Fields},
Phys. Rev. D 81 (2010) 084044 (arXiv:1003.1159);\\
 A.Yu.~Kamenshchik, E.~O. Pozdeeva, A.~Tronconi, G.~Venturi, and S.~Yu. Vernov,
  \textit{Integrable cosmological models with non-minimally coupled scalar fields},
  Class.\ Quant.\ Grav.   \textbf{31} (2014) 105003 (arXiv:1307.1910);\\
 G.~Domenech and M.~Sasaki,
  \textit{Conformal Frame Dependence of Inflation},
  J. Cosmol. Astropart. Phys.  {\bf 1504} (2015)  022
  (arXiv:1501.07699)



\bibitem{Steinwachs:2011zs}
  C.F.~Steinwachs and A.Yu.~Kamenshchik,
  \textit{One-loop divergences for gravity non-minimally coupled to a multiplet of scalar fields: calculation in the Jordan frame. I. The main results},  Phys.\ Rev.\ D {\bf 84} (2011) 024026
  (arXiv:1101.5047)

\bibitem{Kamenshchik:2014waa}
  A.Yu.~Kamenshchik and C.F.~Steinwachs,
  \textit{Question of quantum equivalence between Jordan frame and Einstein frame},
  Phys.\ Rev.\ D {\bf 91} (2015) 084033
  (arXiv:1408.5769)



  \bibitem{Liddle:1994dx}
 A.R.~Liddle, P.~Parsons and J.D.~Barrow,
  \textit{Formalizing the slow roll approximation in inflation},
  Phys.\ Rev.\ D {\bf 50} (1994) 7222
 (arXiv:astro-ph/9408015)
  %%CITATION = ASTRO-PH/9408015;%%


\bibitem{Bamba:2014daa}
  K.~Bamba, Sh.~Nojiri and S.D.~Odintsov,
\textit{Reconstruction of scalar field theories realizing inflation
consistent with the Planck and BICEP2 results},
  Phys. Lett. B \textbf{737} (2014) 374--378 (arXiv:1406.2417)



\bibitem{9} I.L. Buchbinder, S.D. Odintsov and I.M. Lichtzier, \textit{The behaviour of effective coupling constants in 'finite' grand unification theories in curved spacetime}, Class. Quant. Grav. {\bf 6} (1989) 605

\bibitem{4} M. B\"{o}hm and A. Denner, \textit{Features of Finite Quantum Field Theories}, Nucl. Phys. B {\bf 282}
(1987) 206.


 \bibitem{SU2} B.L. Voronov and I.V. Tyutin,
  Yad. Fiz. (Sov. J. Nucl. Phys.) {\bf 23} (1976) 664--675;\\
I.L. Buchbinder and S.D. Odintsov, \textit{Asymptotical Properties Of Nonabelian Gauge Theories In External Gravitational Fields},
Yad. Fiz. \textbf{40} (1984) 1338--1343 (Sov. J. Nucl. Phys. {\bf 40} (1984) 848).

\bibitem{Odintsov7} S.D. Odintsov, D.J. Toms, I.L. Shapiro,
\textit{Asymptotic Freedom Versus Asymptotic Finiteness},
 Int. J. Mod. Phys. A \textbf{6} (1991) 1829--1834

\bibitem{Sami:2012uh}
 M.~Sami, M.~Shahalam, M.~Skugoreva, and A.~Toporensky,
\textit{Cosmological dynamics of nonminimally coupled scalar
field system and its late time cosmic relevance},
Phys. Rev. D \textbf{86} (2012) 103532
(arXiv:1207.6691);\\
 %%CITATION = ARXIV:1207.6691;%%
M.A.~Skugoreva, A.V.~Toporensky, and S.Yu.~Vernov,
\textit{Global stability analysis for cosmological models with
nonminimally coupled scalar fields},
Phys. Rev. D \textbf{90} (2014) 064044  (arXiv:1404.6226)


\bibitem{Gottlober:1993hp}
S.~Gottlober, J.P.~Mucket, and A.A.~Starobinsky,
  \textit{Confrontation of a double inflationary cosmological model with observations},
  Astrophys.\ J.\  {\bf 434} (1994) 417
  (arXiv:astro-ph/9309049)

\bibitem{KLS}
  L.~Kofman, A.D.~Linde and A.A.~Starobinsky,
  \textit{Reheating after inflation},
  Phys.\ Rev.\ Lett.\  {\bf 73} (1994) 3195
  (arXiv:hep-th/9405187)

\bibitem{ReheatingReview}
  B.A.~Bassett, S.~Tsujikawa and D.~Wands,
  \textit{Inflation dynamics and reheating},
  Rev.\ Mod.\ Phys.\  {\bf 78}  (2006) 537
  (arXiv:astro-ph/0507632);\\
  R.~Allahverdi, R.~Brandenberger, F.Y.~Cyr-Racine and A.~Mazumdar,
  \textit{Reheating in Inflationary Cosmology: Theory and Applications},
  Ann.\ Rev.\ Nucl.\ Part.\ Sci.\  {\bf 60} (2010) 27
  (arXiv:1001.2600);\\
   J.L.~Cook, E.~Dimastrogiovanni, D.A.~Easson and L.M.~Krauss,
  \textit{Reheating predictions in single field inflation},
  J. Cosmol. Astropart. Phys.  {\bf 1504} (2015) 047
  (arXiv:1502.04673);\\
  M.A.~Amin, M.P.~Hertzberg, D.I.~Kaiser and J.~Karouby,
 \textit{ Nonperturbative Dynamics Of Reheating After Inflation: A Review},
  Int.\ J.\ Mod.\ Phys.\ D {\bf 24} (2014) 1530003
  (arXiv:1410.3808)



\end{thebibliography}
\end{document}